\documentclass[twocolumn,amssymb,superscriptaddress,aps,prd,floatfix]{revtex4}
\usepackage{amsmath}
\usepackage[dvips]{graphicx}
\usepackage[dvips]{color}
\usepackage[usenames,dvipsnames]{xcolor}
\newlength{\colw}
\newcommand{\err}[2]{\mbox{$\stackrel{\scriptstyle +#1}{\scriptstyle -#2}$}}

\newcommand{\psibar}{\bar{\psi}}
\newcommand{\half}{\frac{1}{2}}
\newcommand{\bra}{\langle}
\newcommand{\ket}{\rangle}
\newcommand{\braket}[1]{\langle#1\rangle}
\newcommand{\tr}{\operatorname{Tr}}
\renewcommand{\Re}{\operatorname{Re}}
\newcommand{\order}{{\cal O}}
\newcommand{\One}{1\kern-4.5pt1}

\newcommand{\eps}{\varepsilon}
\newcommand{\eq}{\varepsilon_q}
\newcommand{\eg}{\varepsilon_g}
\newcommand{\qq}{\bra qq\ket}

\newcommand{\ntr}{N_{\text{traj}}}

\newcommand{\pbp}{\bra\psibar\psi\ket}

\newcommand{\del}[2]{\frac{\partial#1}{\partial#2}}

\newcommand{\cdeconf}{\cite{Hands:2006ve}}
\newcommand{\cquarkyonic}{\cite{Hands:2010gd}}
\newcommand{\cprev}{\cite{Hands:2006ve,Hands:2010gd}}

\begin{document}

\title{Towards the phase diagram of dense two-color matter}

\author{Seamus Cotter}

\affiliation{
Department of Mathematical Physics, National University of Ireland Maynooth,
Maynooth, County Kildare, Ireland.}

\author{Pietro Giudice}

\affiliation{
       Department of Physics, College of Science, Swansea University,
       Singleton Park, Swansea SA2 8PP, U.K.
       }
\author{Simon Hands}

\affiliation{
       Department of Physics, College of Science, Swansea University,
       Singleton Park, Swansea SA2 8PP, U.K.
       }

\author{Jon-Ivar Skullerud}

\affiliation{
Department of Mathematical Physics, National University of Ireland Maynooth,
Maynooth, County Kildare, Ireland.}

\affiliation{
Institute for Nuclear Theory, University of Washington, Seattle, WA
98195--1550, USA.}

\begin{abstract}
We study two-color QCD with two flavors of Wilson fermion as a
function of quark chemical potential $\mu$ and temperature $T$.  We find
evidence of a superfluid phase at intermediate $\mu$ and low $T$ where
the quark number density and diquark condensate are both very well
described by a Fermi sphere of nearly-free quarks disrupted by a BCS
condensate. Our results suggest that the quark contribution to
the energy density is negative (and balanced by a positive gauge
contribution), although this result is highly 
sensitive to details of the energy renormalisation.
We also find evidence that the chiral condensate in
this region vanishes in the massless limit.  This region gives way to
a region of deconfined quark matter at 
higher $T$ and $\mu$, with the deconfinement temperature, determined
from the renormalised Polyakov loop, decreasing
only very slowly with increasing chemical potential.  The quark
number susceptibility $\chi_q$ does not exhibit any qualitative change at the
deconfinement transition.  We argue that this is because $\chi_q$ is
not an appropriate measure of deconfinement for 2-color QCD at high
density.
\end{abstract}

\pacs{11.15.Ha,12.38.Aw,21.65.Qr}
         
\maketitle

\section{Introduction}

Despite over a decade of intensive efforts to unveil the phase
structure of strongly interacting matter at high density (beyond a few
times the nuclear saturation density) and low temperature, even the
question of which phases exist remains unanswered.  A quantitative
knowledge of this region would allow us to answer many questions
regarding the structure and properties of compact stars, including the
question of whether deconfined quark matter can exist inside such
stars.  The reason for the lack of definite progress on this issue is
that standard weak-coupling methods are inapplicable except at
asymptotically high densities, while the various model approaches that
have been employed have not been sufficiently constrained by input
from experiment or first-principles theoretical calculations to yield
reliable information in the region of interest.  Thus, while a wealth
of information exists regarding possible phases and their properties
in various models, no reliable, quantitative results are available as
yet.  For a recent review of high-density QCD, see
  Ref.~\cite{Fukushima:2010bq}.

Many of the outstanding questions could in principle be answered by
lattice QCD simulations, but these have been hindered by the notorious
sign problem.  While no method has as yet been shown to solve the sign
problem for QCD, lattice simulations may still constrain model
calculations by providing first-principles, nonperturbative results
for QCD-like theories without a sign problem.  This is the main aim of
the present study.

Among these theories, QCD with gauge group SU(2) (two-color QCD or
QC$_2$D) is of particular interest in that it shares most of the
salient features of real QCD (eg, confinement, dynamical chiral
symmetry breaking and long-range interactions).  It differs from QCD
in that the baryons of the theory are bosons, and the lightest baryon
is a pseudo-Goldstone boson, degenerate with the pion (note though, that 
SU(2) models with adjoint matter \cite{Hands:2000ei} and G$_2$ with fundamental
matter \cite{Maas:2012wr}, both of which are free from a sign
problem, are expected
to contain fermionic baryons in the physical spectrum).  Therefore,
instead of a normal nuclear matter phase this theory has a
superfluid state characterised by condensation of these
baryons, which at this point become true Goldstone bosons.
This has been observed in a number of 
lattice simulations; in particular, the excitation spectrum
  including the Goldstone bosons has been studied in
  Refs.~\cite{Kogut:2001na,Hands:2007uc}.
A transition to a state of deconfined
quark matter is expected at high chemical potential $\mu$ (see however
\cite{Brauner:2009gu}), and evidence of this was found in \cprev.  The
precise nature of this transition remained unclear, however, and in
this paper we will attempt to answer some of the outstanding questions
about this.

An intriguing possibility is that in an intermediate r\'egime, strongly
interacting matter may enter a chirally symmetric and confined phase,
dubbed {\em quarkyonic} \cite{McLerran:2007qj}.  In \cquarkyonic, it
was suggested that the scaling of thermodynamic quantities with $\mu$
in the intermediate r\'egime could be a sign of such a phase.  It was
not possible to draw any further conclusions, not least because the
presence of a non-zero diquark source $j\not=0$, introduced to stabilise the
simulations, distorted the $\mu$-dependence of the relevant
quantities.  This will be remedied in the present paper.

This paper is organised as follows.  In Section~\ref{sec:vacuum} we
present results from simulations at zero chemical
potential.  These results allow us to map out lines of constant
physics, including the line of zero quark mass, which will in the
future allow us to perform controlled extrapolations to the continuum
and chiral limits, and also by varying $N_\tau$ at fixed cutoff to estimate the
critical temperature $T_d$ for deconfinement at $\mu=0$.  In addition, these results form a large part of
the input into the renormalisation of energy densities, which is
described and carried out in Section~\ref{sec:karsch}.
Section~\ref{sec:results} contains the bulk of our results for the
$(\mu,T)$ phase diagram.  After addressing some general technical
issues in Section~\ref{sec:technical}, we present in Section~\ref{sec:phases}
results for the order parameters for superfluidity and deconfinement,
giving us an outline of the $(\mu,T)$ phase diagram.
Section~\ref{sec:thermo} contains results for the thermodynamic
quantities, baryon density and (renormalised) energy density, while
Section~\ref{sec:susc} contains results for the quark number
susceptibility (preliminary
results from this work were presented in Ref.~\cite{Giudice:2011zu}),
and in Section~\ref{sec:chisb} we investigate chiral
symmetry breaking and restoration.  Finally, in
Section~\ref{sec:conclude} we summarise our results and their
implications.

\section{Simulation details and vacuum phase structure}
\label{sec:vacuum}

We study QC$_2$D with a conventional Wilson action for the gauge
fields and two flavours of Wilson fermion.  The fermion action is
augmented by a gauge- and iso-singlet diquark source term which serves
the dual purpose of lifting the low-lying eigenvalues of the Dirac
operator and allowing a controlled study of diquark condensation.  The
quark action is
\begin{equation}
S_Q+S_J=\sum_{i=1,2}\bar\psi_iM\psi_i
 + \kappa j[\psi_2^{tr}(C\gamma_5)\tau_2\psi_1-h.c.],
\label{eq:Slatt}
\end{equation}
with
\begin{align}
M_{xy}=\delta_{xy}-\kappa\sum_\nu&\Bigl[(1-\gamma_\nu)e^{\mu\delta_{\nu0}}U_\nu(x)\delta_{y,x+\hat\nu}\nonumber\\
&+(1+\gamma_\nu)e^{-\mu\delta_{\nu0}}U^\dagger_\nu(y)\delta_{y,x-\hat\nu}\Bigr].
\label{eq:Mwils}
\end{align}
Further details about the action and the Hybrid Monte Carlo algorithm
used can be found in \cite{Hands:2006ve}.

\begin{table}
\begin{tabular*}{\colw}{l@{\extracolsep{\fill}}lllrlll}
\hline
$\beta$ & $\kappa$ & $N_s$ & $N_\tau$ & $N_{traj}$ & $am_\pi$&
$m_\pi/m_\rho$ & $a$ (fm) \\ \hline
1.7\ \ & 0.1780 & 12 & 24 & 500 & 0.779(7) & 0.804(10) & 0.229(3) \\
1.7 & 0.1790 & 12 & 24 & 1050 & 0.683(5) & 0.783(12) &  0.213(8) \\
1.7 & 0.1810 & 12 & 24 & 500 & 0.438(15) & 0.61(5) &  0.189(4) \\
1.8 & 0.1740 & 12 & 24 & 2000 & 0.640(4) & 0.778(7) & 0.178(8) \\
1.8 & 0.1750 & 12 & 24 &  880 & 0.490(9) & 0.67(2)  & 0.174(8) \\
1.9 & 0.1680 & 12 & 24 & 1570 & 0.645(8) & 0.805(9) & 0.178(6) \\
1.9 & 0.1685 & 12 & 24 & 2000 & 0.589(4) & 0.780(9) & 0.153(18) \\
1.9 & 0.1690 & 12 & 24 & 1000 & 0.517(11) & 0.71(2) & 0.144(8) \\
2.0 & 0.1620 & 12 & 24 & 1000 & 0.638(7) & 0.830(9) & 0.164(5) \\
2.0 & 0.1625 & 16 & 32 & 2000 & 0.586(3) & 0.820(8) & \\
2.0 & 0.1627 & 16 & 32 & 2000 & 0.562(4) & 0.809(8) & \\
2.0 & 0.1630 & 12 & 24 & 1000 & 0.524(10) & 0.758(16) & 0.145(3) \\
    &        & 16 & 32 & 2000 & 0.508(4) & 0.785(9) & \\
2.1 & 0.1570 & 16 & 32 & 1600 & 0.536(3) & 0.836(8) & \\
2.1 & 0.1580 & 16 & 32 & 2100 & 0.405(5) & 0.770(12) & \\ \hline
\end{tabular*}
\caption{Simulation parameters, pi and rho masses and
  lattice spacing at $\mu=j=0$.}
\label{tab:params-mu0}
\end{table}

We have performed an extensive exploration of the parameter space in
the vacuum ($T=\mu=j=0$) in the range $\beta=1.7-2.1$.  The parameters
used are shown in Table~\ref{tab:params-mu0}, together with the values
obtained for the pion (pseudoscalar meson) mass $m_\pi$, ratio of pion
to rho (vector meson) mass $m_\pi/m_\rho$ and lattice spacing $a$.
The lattice spacing was determined by fitting the static quark
potential to the Cornell form $V(r)=C+\alpha/r+\sigma r$ and taking
the string tension to be $\sqrt{\sigma}=440$MeV. 

\begin{table}
\begin{center}
\begin{tabular}{r|lllll}\hline
$\beta$ & 1.7 & 1.8 & 1.9 & 2.0 & 2.1 \\\hline
$\kappa_c$ & 0.18226\err{8}{8} & 0.17644\err{15}{11}
 & 0.17089\err{20}{19} & 0.16456\err{14}{10} & 0.15935\err{8}{8}
\\ \hline
\end{tabular}
\end{center}
\caption{Critical hopping parameter $\kappa_c$ given by
  $m_\pi^2(\kappa_c)=0$, for different values of $\beta$.}
\label{tab:chiral}
\end{table}

We can determine the value $\kappa_c(\beta)$ where the quark mass
vanishes by performing a linear extrapolation of $m_\pi^2$ in
$1/\kappa$ for each value of $\beta$.  The results of this are shown
in Table~\ref{tab:chiral}.

\begin{figure}[tb]
\includegraphics*[width=\colw]{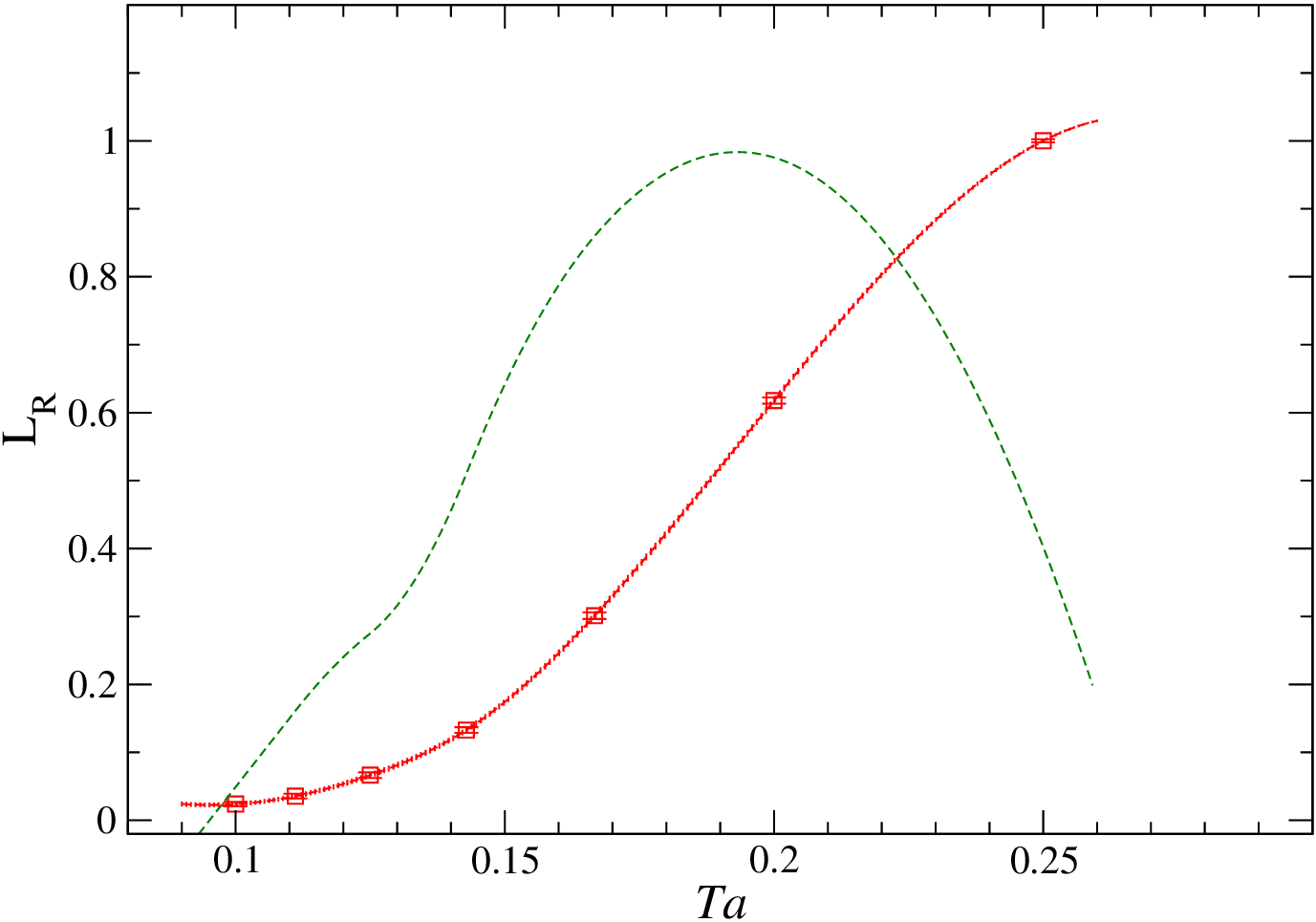}
\caption{The renormalised Polyakov loop $L_R$ as a function of
  temperature $T$, for $16^3\times N_\tau$ lattices at $\beta=1.9,
  \kappa=0.168, \mu=j=0$.  The red (solid) band is a cubic spline
  interpolation between the data points, and the green (dashed) curve shows the
  derivative of the interpolation curve, divided by a factor of 10.}
\label{fig:poly-Tscan}
\end{figure}

We have also investigated the thermal deconfinement transition at
$\mu=0$ using the fixed-scale approach.  We have generated
configurations with $N_\tau=4-10$ at $\beta=1.9,\kappa=0.168$,
corresponding to a temperature range of 113--281 MeV.  At each
temperature we have computed the Polyakov loop $\bra L\ket$, which is
an order parameter for deconfinement of static color charges in the
pure gauge theory, and exhibits a rapid crossover in a theory with
dynamical fermions.  It is related to the free energy $F_q$ of a
static quark by
\begin{equation}
L = e^{-F_q(T)/T}\,.
\end{equation}
The free energy $F_q$ is only defined up to an additive
renormalisation constant $\Delta F$, which depends on the bare
couplings $\beta, \kappa$.  Different prescriptions for determining
this constant correspond to different renormalisation schemes.  We
have imposed the condition that the renormalised Polyakov loop on our
$N_\tau=4$ lattice ($T=263$ MeV) is equal to 1, or in other words,
the free energy is zero at this temperature.  We can then
compute the renormalised Polyakov loop $L_R(T)$ at any other
temperature $T$ from the bare Polyakov loop $L_0$ via
\begin{equation}
\begin{split}
L_R(T) &= e^{-F_R(T)/T} = e^{-(F_0(T)+\Delta F)/T}\\
 &= L_0(T)e^{-\Delta F/T} = Z_L^{N_\tau}L_0(T=1/aN_\tau)\,,
\end{split}
\label{eq:polyakov-renorm}
\end{equation}
where $Z_L=\exp(-a\Delta F)=L_0(N_\tau=4)^{-1/4}$ (this
procedure was first outlined in Ref.~\cite{Borsanyi:2012xf}).  The
results are shown in Fig.~\ref{fig:poly-Tscan}, as a function of
$aT=1/N_\tau$.
The red (solid) curve in Fig.~\ref{fig:poly-Tscan} is a cubic spline
interpolation between the data points.  Taking the derivative of this
(denoted by the green, dashed curve), we find the maximum at $Ta=0.193$.  If
we instead use an Akima spline to interpolate, the maximum of the
derivative appears at $Ta=0.183$.  Taking the cubic spline as our best
estimate and conservatively estimating the uncertainty to be twice the
difference between the Akima and cubic spline estimates, our result
for the deconfinement temperature is $T_d(\mu=0)$ is $T_da=0.193(20)$ or
$T_d=217(23)$ MeV.

\section{Renormalisation of energy densities}
\label{sec:karsch}

To determine the energy density, it is convenient to introduce
different lattice spacings $a_s, a_\tau$ in the space and time
directions, with an anisotropy parameter $\xi\equiv a_s/a_\tau$. 
The energy density $\varepsilon$ is then given by \cite[sec.~5.4.1]{M&M}
\begin{equation}
\eps(T)=-\frac{1}{V}{\del{Z}{T^{-1}}}\biggr\vert_{V}
  =-\frac{\xi}{N_s^3N_\tau a_s^3a_\tau}
  \bigg\bra\del{S}{\xi}\bigg\vert_{a_s}\bigg\ket\,,
\label{eq:energy}
\end{equation}
where we have used $V=(N_sa_s)^3$, $T^{-1}=N_\tau a_\tau$, and
\begin{equation}
\del{}{a_\tau}\bigg\vert_{a_s}
 =-\frac{a_s}{a_\tau^2}\del{}{\xi}\bigg\vert_{a_s}.
\end{equation}
The partial derivatives must be taken with all other physical
parameters kept fixed.  In our case, this means that the physical
quark mass, and therefore the ratio $m_\pi/m_\rho$, is kept fixed.

The anisotropic action $S=S_G+S_Q+S_J$ describing $N_f=2$ Wilson quark
flavors is given by
\begin{align}
S_G =& -\frac{\beta}{N_c}\left[
\frac{1}{\gamma_{g}}\sum_{x,i<j} \Re\tr U_{ij}(x)+
\gamma_g\sum_{xi}\Re\tr U_{i0}(x)\right]\,,\\
S_Q = &
\sum_{x,\alpha}\bigg[\psibar^\alpha(x)\psi^\alpha(x)
 + \gamma_q\kappa\psibar^\alpha(x)(D_0\psi)^\alpha(x)\bigg]\,,
\notag\\
& +\kappa\sum_{x,\alpha,i}\psibar^\alpha(x)(D_{i}\psi)^\alpha(x)\\
S_J =&  \kappa j\sum_x[\psi^{2tr}(x)C\gamma_5\tau_2\psi^1(x)
-\bar\psi^1(x)C\gamma_5\tau_2\bar\psi^{2tr}(x)]\,,
\end{align}
with
\begin{align}
(D_i\psi)^\alpha(x) =&\,
(\gamma_i-1)U_i(x)\psi^\alpha(x+\hat\imath)\notag\\
&-(\gamma_i+1)U_i^\dagger(x-\hat\imath)\psi^\alpha(x-\hat\imath)\,,\\
(D_0\psi)^\alpha(x) =&\,
(\gamma_0-1)U_0(x)e^\mu\psi^\alpha(x+\hat0)\notag\\
&-(\gamma_0+1)U_0^\dagger(x-\hat0)e^{-\mu}\psi^\alpha(x-\hat0)\,.
\end{align}
We also define
\begin{equation}
\beta_s=\frac{\beta}{\gamma_g};\quad
\beta_t=\gamma_g\beta;\quad
\kappa_t=\gamma_q\kappa_s=\gamma_q\kappa\,.
\end{equation}
The parameters $\gamma_g$ and $\gamma_q$ are the bare gluon and quark
anisotropies, which in our formalism will be taken to be independent.

Substituting these expressions into \eqref{eq:energy} (and dropping
the $\vert_{a_s}$ from all partial derivatives as it will be
understood), we then readily derive
\begin{align}
\frac{\eg}{T^4} =&
 -\xi\left(\frac{N_\tau a_\tau}{N_sa_s}\right)^3
  \bigg\bra\del{S_G}{\xi}\bigg\ket\notag\\
 =& \frac{3N_\tau^4}{\xi^2{N_c}}\biggl[
\bra\Re\tr U_{ij}\ket
 \left(\gamma_g^{-1}{\del{\beta}{\xi}}
   +\beta{\del{\gamma_g^{-1}}{\xi}}\right)\notag\\
 & \phantom{\frac{3N_\tau^4}{\xi^2}}+
 \bra\Re\tr U_{i0}\ket
 \left(\gamma_g\del{\beta}{\xi}+\beta\del{\gamma_g}{\xi}\right)\biggr].
\label{eq:epsG}
\end{align}
This coincides with the first part of Eq.~(17) of
Ref.~\cite{Levkova:2006gn}.  The terms in angled brackets are the
average spatial and temporal plaquettes respectively, and the terms
multiplying them are what are usually known as the Karsch
coefficients. In the weak coupling isotropic limit $\beta\to\infty,
\gamma_g=1$ we have
\begin{equation}
\del{\gamma_g}{\xi}=-\del{\gamma_g^{-1}}{\xi}=1;
\quad \del{\beta}{\xi}=-a\del{\beta}{a}=0, 
\end{equation}
and we recover the expression used in \cprev:
\begin{equation}
\frac{\eg^0}{T^4}
 =\frac{3N_\tau^4\beta}{N_c}\left[\bra\Re\tr U_{i0}\ket-\bra\Re\tr U_{ij}\ket\right].
\end{equation}

The quark contribution to the energy density is given by
\begin{align}
\frac{\eq}{T^4} &= -\xi\left(\frac{N_\tau a_\tau}{N_sa_s}\right)^3
 \biggl\bra\del{S_Q}{\xi}\biggr\ket\label{eq:epsQ}\\
&= -\frac{N_\tau^4}{\xi^2}\biggl[\big\bra\sum_i\psibar D_i\psi\big\ket
  \del{\kappa}{\xi}+ \big\bra\psibar D_0\psi\big\ket
  \left(\gamma_q\del{\kappa}{\xi}
   +\kappa\del{\gamma_q}{\xi}\right)\biggr].\notag
\end{align}
The terms in angled brackets are calculated
using a stochastic estimator. Note a potentially useful identity
\begin{equation}
\gamma_q\kappa\bra\bar\psi D_0\psi\ket+\kappa\sum_i\bra\bar\psi D_i\psi\ket
+\bra\bar\psi\psi\ket=-\tr1=-4N_cN_f.
\end{equation}
Note that we have taken explicit account of the minus sign associated
with closed fermion loops in the definition of the bilinear expectation values, 
ie. $\bra\bar\psi\Gamma\psi\ket\equiv-\tr(\Gamma M^{-1}).$
It is therefore sufficient to evaluate the first and third terms on
the LHS, enabling the second term, which enters into Eq.~\eqref{eq:epsQ},
to be estimated.  In the isotropic limit $\gamma_q=\xi=1$ this
  reduces to
\begin{equation}
\frac{\eq}{T^4}
 = N_\tau^4\Big[(4N_fN_c+\pbp)\kappa^{-1}\del{\kappa}{\xi}
- \kappa\del{\gamma_q}{\xi}\bra\psibar D_0\psi\ket\Big]\,.
\end{equation}
In the weak coupling isotropic limit $\partial\kappa/\partial\xi=0,
\partial\gamma_q/\partial\xi=1$ and we
recover
\begin{equation}
\frac{\eq^0}{T^4}=-N_\tau^4\kappa\bra\bar\psi D_0\psi\ket\,,
\end{equation}
which coincides up to an overall sign with the expression in \cprev,
where the fermion's  Grassmann nature was ignored.

Finally, the diquark contribution is given by
\begin{align}
\frac{\eps_J}{T^4}
 &= \frac{N_\tau^4}{\xi^2}\left(\del{(\kappa j)}{\xi}\right)
 \bra-\bar\psi^1C\gamma_5\tau_2\bar\psi^{2tr}
 +\psi^{2tr}C\gamma_5\tau_2\psi^1\ket\notag\\
 &= \frac{2N_\tau^4}{\xi^2}
 \left(\del{j}{\xi}+\frac{j}{\kappa}\del{\kappa}{\xi}\right)
 \bra qq\ket
\end{align}
in the notation of \cite{Hands:2006ve}.  However, in the U(1)$_B$-symmetric
limit $j\to0$ the second term inside the brackets vanishes, and since
this limit is always found at $j=0$ for any anisotropy $\xi$, the
first term also vanishes here.

Similarly, the trace anomaly is given by
\begin{equation}
T_{\mu\mu} \equiv \eps-3p
 = \frac{T}{V}\Big\bra a_s\del{S}{a_s}\big|_\xi\Big\ket\,.
\label{eq:Tmumu}
\end{equation}
With our anisotropic action the quark and gluon contributions are
given by
\begin{align}
(T_{\mu\mu})_g &=
\frac{3}{\xi^2{N_c}}\biggl[
\bra\Re\tr U_{ij}\ket
 \left(\gamma_g^{-1}a\del{\beta}{a}
   +\beta a\del{\gamma_g^{-1}}{a}\right)\notag\\
 & \phantom{=\frac{3}{\xi^2N_c}}+
 \bra\Re\tr U_{i0}\ket
 \left(\gamma_ga\del{\beta}{a}+\beta a\del{\gamma_g}{a}\right)\biggr]\,,
\label{eq:Tg-xi}\\
(T_{\mu\mu})_q &=
\frac{1}{\xi^2}\biggl[\big\bra\sum_i\psibar D_i\psi\big\ket
  a\del{\kappa}{a} \notag\\
& \phantom{=\frac{\kappa}{\xi^2}}+ \big\bra\psibar D_0\psi\big\ket
  \left(\gamma_qa\del{\kappa}{a}
   +\kappa a\del{\gamma_q}{a}\right)\biggr]\,.
\label{eq:Tq-xi}
\end{align}
However, in the isotropic limit, the bare anisotropies are always 1,
and hence the derivatives $\partial\gamma_{g,q}/\partial a$ vanish.
We are then left with the standard expressions for the trace anomaly,
\begin{align}
(T_{\mu\mu})_g &=
-a\del{\beta}{a}\frac{3}{N_c}\bra\Re\tr U_{ij}+\Re\tr U_{i0}\ket\,,
\label{eq:Tg}\\
(T_{\mu\mu})_q &=-
  a\del{\kappa}{a}\kappa^{-1}(4N_fN_c+\pbp)\,.
\label{eq:Tq}
\end{align}
Eqs.~(\ref{eq:Tg},\ref{eq:Tq}) differ from the expressions used in \cprev\
by an overall factor $\beta$ and an overall sign respectively; the
resulting error is corrected in this paper.

So, in order to evaluate the full energy density (ignoring $j\neq0$) from
Eqs. (\ref{eq:epsG},\ref{eq:epsQ}) we need the following, which go into the
definition of the ``Karsch coefficients'':
\begin{equation}
\del{\beta}{\xi}\,;\quad \del{\gamma_g}{\xi}\,;\quad
\del{\kappa}{\xi}\,;\quad \del{\gamma_q}{\xi}\,.
\label{eq:Karsch}
\end{equation}
These are computed using the method presented in
\cite{Levkova:2006gn,Morrin:2009}.  In addition to the bare anisotropies we
define the physical anisotropies $\xi_g=a_s/a_\tau$ as determined from
gluonic observables such as the ``sideways potential'' \cite{Klassen:1998ua}, and
$\xi_q=a_s/a_\tau$ as determined from a meson dispersion relation.  For a
parameter set corresponding to a {\it physical\/} system $\xi_g$ and
$\xi_q$ should be equal, since otherwise a massless meson would not
propagate at the correct speed of light; choosing the bare parameters
to bring this about is a non-trivial tuning problem
\cite{Levkova:2006gn,Morrin:2006tf}. In attempting to calculate
the Karsch coefficients for the parameter set $\beta=1.9$,
$\kappa=0.168$, we do not attempt this
tuning, but rather simulate unphysical ensembles with either
$\gamma_g$ or $\gamma_q$ set to unity; the parameters are given in
Table~\ref{tab:aniso-params}.  In addition we use the isotropic
ensembles given in Table~\ref{tab:params-mu0}.

\begin{table*}
\begin{tabular}{rrrrrr|llll}
$\beta_s$ & $\beta_t$ & $\kappa_s$ & $\kappa_t$ & $\gamma_g$ &
  $\gamma_q$ & $\xi_g$ & $\xi_q$ & $m_\pi/m_\rho$ & $a_s$(fm)\\ \hline
1.90 & 1.90 & 0.1680 & 0.1680 & 1.0 & 1.0
 & 0.968\err{2}{2} & 1.07\err{2}{3} & 0.807\err{5}{5} & 0.178\err{4}{6} \\\hline
2.37 & 1.52 & 0.168 & 0.168 & 0.8 & 1.0
 & 0.720\err{2}{2} & 0.853\err{14}{10} & 0.805\err{4}{5} & 0.177\err{4}{3} \\
1.27 & 2.83 & 0.168 & 0.168 & 1.5 & 1.0
 & 1.321\err{5}{5} & 1.32\err{4}{3} & 0.648\err{8}{12} & 0.125\err{3}{6} \\
1.90 & 1.90 & 0.180 & 0.157 & 1.0 & 0.87
 & 0.747\err{4}{4} & 0.78\err{4}{3} & 0.746\err{21}{13} & 0.107\err{3}{6} \\ 
1.90 & 1.90 & 0.147 & 0.192 & 1.0 & 1.3
 & 1.146\err{4}{4} & 1.53\err{2}{2} & 0.946\err{1}{1} & 0.229\err{7}{13} \\ \hline
1.80 & 1.80 & 0.1740 & 0.1740 & 1.0 & 1.0
 & 0.989\err{3}{3} & 1.03\err{1}{3} & 0.777\err{6}{8} & 0.177\err{6}{8} \\
1.90 & 1.90 & 0.1685 & 0.1685 & 1.0 & 1.0 
 & 0.945\err{5}{6} & 0.98\err{3}{3} & 0.760\err{10}{18} & 0.153\err{7}{18} \\
2.00 & 2.00 & 0.1620 & 0.1620 & 1.0 & 1.0
 & 0.921\err{4}{5} & 0.99\err{3}{4} & 0.829\err{9}{9} & 0.166\err{1}{3} \\
2.00 & 2.00 & 0.1630 & 0.1630 & 1.0 & 1.0
 & 0.881\err{5}{5} & 1.04\err{5}{4} & 0.773\err{11}{11} & 0.148\err{2}{1} \\ \hline
\end{tabular}
\caption{Anisotropic lattice parameters and anisotropy
  results. The uncertainties are purely statistical. 
}
\label{tab:aniso-params}
\end{table*}

For each ensemble we compute the ratio $M=(m_\pi/m_\rho)^2$, the
lattice spacing $a\equiv a_s$, the gluon anisotropy $\xi_g$ (from the sideways
potential) and the quark anisotropy $\xi_q$ (from the pion dispersion
relation).  The quark and gluon anisotropies are combined to form the
average anisotropy $\xi_+=\half(\xi_g+\xi_q)$ and the anisotropy
mismatch $\xi_-=\xi_g-\xi_q$.  Each of these quantities is fitted to a
linear function in the bare parameters,
\begin{align}
 \xi_+-1 &= a_1\Delta\gamma_g + b_1\Delta\gamma_q
 + c_1\Delta\beta + d_1\Delta\kappa\,,\label{eq:xifit}\\
 \frac{a-a_0}{a_0} &= a_2\Delta\gamma_g + b_2\Delta\gamma_q
 + c_2\Delta\beta + d_2\Delta\kappa\,,\label{eq:afit}\\
 \frac{M-M_0}{M_0} &= a_3\Delta\gamma_g + b_3\Delta\gamma_q
 + c_3\Delta\beta + d_3\Delta\kappa\,,\label{eq:Mfit}\\
 \xi_- &= a_4\Delta\gamma_g + b_4\Delta\gamma_q
 + c_4\Delta\beta + d_4\Delta\kappa\,,\label{eq:ximfit}
\end{align}
where $a_0$ and $M_0$ are the values of $a$ and $M$ at the reference
point $\beta=1.9,\kappa=0.168,\gamma_g=\gamma_q=1$, and $\Delta x$ is
the deviation of the bare parameter $x$ from its value at the same
reference point.  Inverting the $4\times4$ matrix of coefficients
$(a_i,b_i,c_i,d_i)$ gives us the ``generalised Karsch coefficients'',
which are the derivatives of the bare parameters with respect to the
``physical'' parameters~\footnote{Strictly speaking, $\xi_-$ is not a
  physical parameter since it denotes the deviation from the physical
  condition $\xi_q=\xi_g$.  However, including $\xi_-$ means that the
  other derivatives are taken at fixed $\xi_-=0$, ie on the physical
  surface, and $\xi_+$ is the physical anisotropy $\xi$.} $\xi_+,
\xi_-, a, M$.  The first column gives us the Karsch coefficients
\eqref{eq:Karsch}, while the second column gives us the beta functions
$\partial\beta/\partial a, \partial\kappa/\partial a$.

Since we do not need to renormalise the pressure, knowledge of the
beta-functions is not required here.  However, we can use information
about them to perform consistency checks.  In the isotropic limit, two
of the Karsch coefficients can be expressed in terms of
beta-functions, since
\begin{equation}
\del{\beta}{\xi}\bigg\vert_{\xi=1}=-a\del{\beta}{a}\,;\quad
 \del{\kappa}{\xi}\bigg\vert_{\xi=1}=-a\del{\kappa}{a}\,.
\label{eq:karsch-beta}
\end{equation}
We can also independently estimate the beta-functions from the
isotropic results in Sec.~\ref{sec:vacuum}, by taking derivatives wrt
$a$ along lines of constant physics.

\begin{figure}[tb]
\includegraphics*[width=\colw]{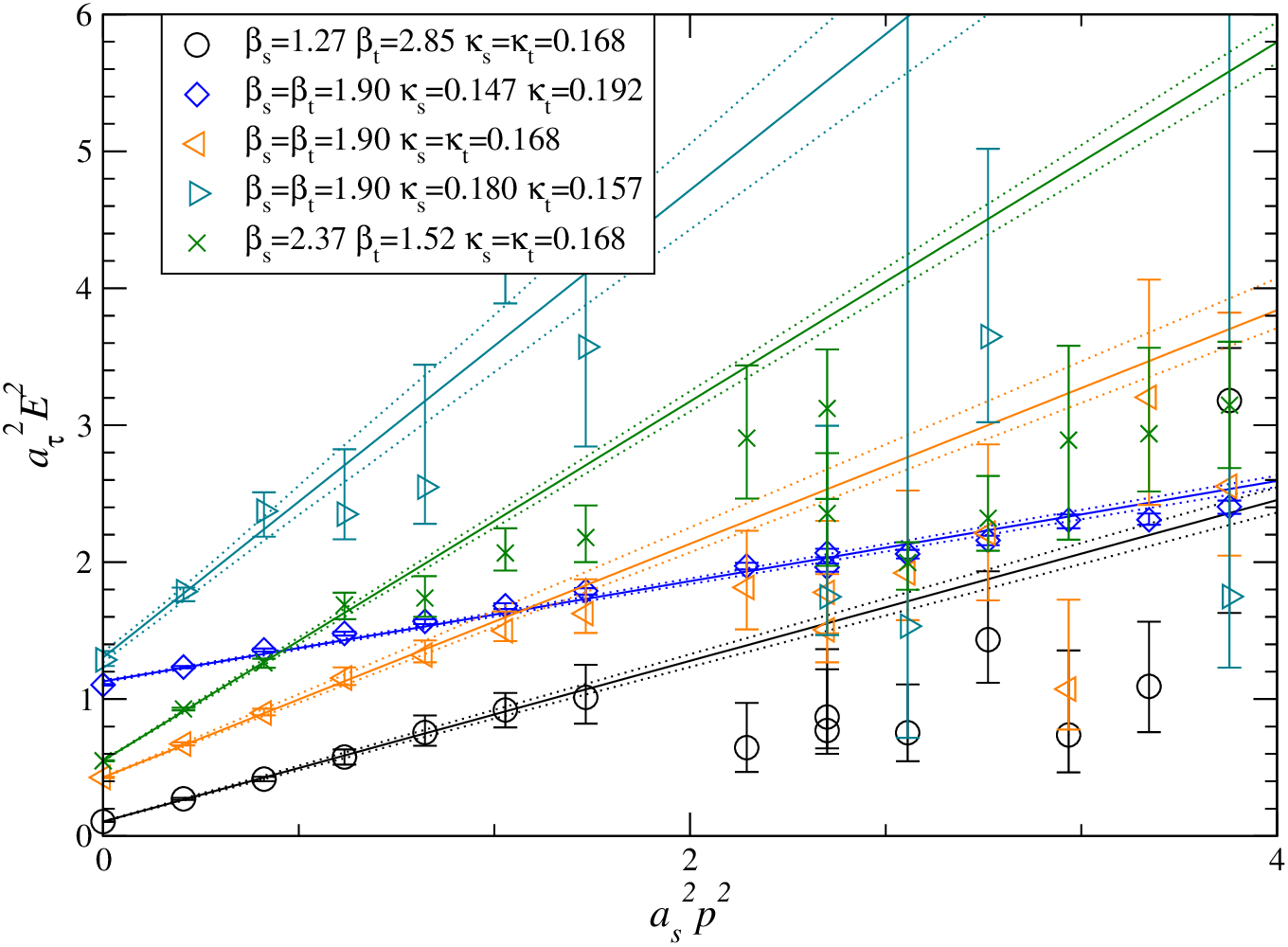}
\caption{The pion dispersion relation from the anisotropic
  $12^3\times24$ lattices in Table~\ref{tab:aniso-params}.}
\label{fig:dispersion}
\end{figure}

\begin{figure}[tb]
\includegraphics*[width=\colw]{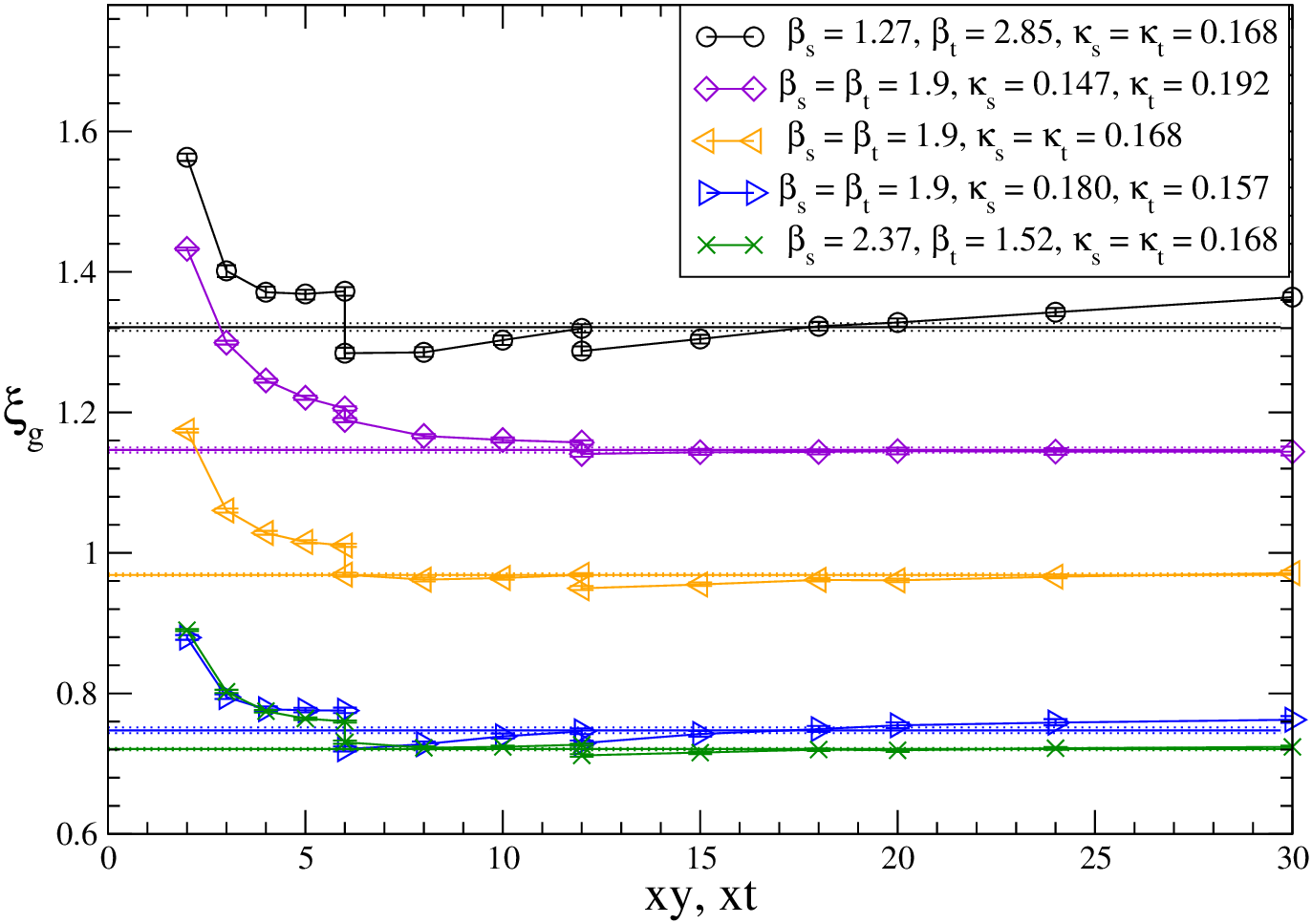}
\caption{The gauge anisotropy from the anisotropic
  $12^3\times24$ lattices in Table~\ref{tab:aniso-params}, computed
  according to Eq.~\eqref{eq:gauge-aniso}.}
\label{fig:sideways}
\end{figure}

Results for the observables on our anisotropic lattices as well as the
isotropic lattices used in this study, are given in
Table~\ref{tab:aniso-params}.  Figs~\ref{fig:dispersion} and
\ref{fig:sideways} illustrate the determination of the quark and gluon
anisotropies respectively.  The gluon anisotropy in
Fig.~\ref{fig:sideways} was computed using \cite{Loan:2003hv}
\begin{equation}
\xi_g = \frac{V_{xt}(R_2)-V_{xt}(R_1)}{V_{xy}(R_2)-V_{xy}(R_1)}\,,
\label{eq:gauge-aniso}
\end{equation}
where $V_{xt}(x),V_{xy}(x)$ are the potentials obtained from Wilson loops in
the $(x,t)$ and $(x,y)$ plane respectively,
\begin{align}
W_{ss}(x,y) &\sim Z_{xy}e^{-yV_{xy}(x)}\,,
W_{st}(x,t) &\sim Z_{xt}e^{-tV_{xt}(x)}\,,
\end{align}
which is valid for large $x$ and $t,y$.  The fermion anisotropy is
determined from the pion dispersion relation,
\begin{equation}
a_\tau^2E^2 = a_\tau^2m_\pi^2 + \frac{a_s^2p^2}{\xi_q^2}\,.
\label{eq:dispersion}
\end{equation}
Hence, a straight-line fit of $a_\tau^2E^2$ vs $a_s^2p^2$, as shown in
Fig.~\ref{fig:dispersion}, will give the anisotropy $\xi_q$.

\begin{table}[htb]
\begin{tabular}{|cc|rrrr|r|}\hline
 & $i$ & $a_i$ & $b_i$ & $c_i$ & $d_i$ & $\chi^2/N_{df}$ \\\hline
$\xi_+$ & 1 & 0.761\err{30}{29} & -1.66\err{0.88}{0.49} & -2.58\err{0.70}{0.39}
 & -39\err{12}{7} & 19.4\\
$a$ & 2 & -0.503\err{48}{55} & -5.14\err{0.79}{1.42} & -5.94\err{0.74}{1.01}
 & -88\err{10}{17} & 2.8 \\
$M$ & 3 & -0.531\err{29}{46} & 0.15\err{1.10}{0.57} & -0.59\err{96}{61}
 & -15\err{16}{8} & 22.6 \\
$\xi_-$ & 4 & 0.096\err{28}{29} & -0.84\err{52}{85} & -0.46\err{38}{73}
 & -6\err{7}{12} & 1.9 \\ \hline
\end{tabular}
\caption{Results for the fits to
  Eqs.~\eqref{eq:xifit}--\eqref{eq:ximfit}.  $\chi^2/N_{df}$ is the
  $\chi^2$ per degree of freedom for each fit.}
\label{tab:anisofits}
\end{table}

\begin{table}[htb]
\begin{tabular}{|c|rrrr|}\hline
$c_i$ & $\del{c_i}{\xi_+}$ & $a\del{c_i}{a}$ & $M\del{c_i}{M}$
 & $\del{c_i}{\xi_-}$ \\\hline
$\gamma_g$ & 0.90\err{4}{14} & -0.51\err{19}{10} & 0.13\err{32}{58}
 & 1.4\err{1.2}{1.6} \\
$\gamma_q$ & 0.13\err{40}{5} & 0.22\err{12}{70} & -0.55\err{2.11}{0.29}
 & -2.9\err{5.7}{0.6} \\
$\beta$ & 0.59\err{0.24}{1.37} & -1.4\err{2.3}{0.5} & 3.7\err{1.9}{7.0}
 & 8\err{8}{19} \\
$\kappa$ & -0.052\err{69}{15} & 0.075\err{24}{99} & -0.22\err{35}{8}
 & -0.39\err{88}{23} \\ \hline
\end{tabular}
\caption{Results for the generalised Karsch coefficients $\partial
  c_i/\partial x_i$.  The numbers in the first column are the actual
  Karsch coefficients, while the second column gives the beta functions.} 
\label{tab:karsch}.
\end{table}

\begin{table}[htb]
\begin{tabular}{|c|rr|}\hline
$c_i$ & $a\del{c_i}{a}$ & $M\del{c_i}{M}$ \\\hline
$\beta$ & -1.02\err{17}{29} & 0.73\err{26}{13} \\
$\kappa$ & 0.057\err{15}{9} & -0.047\err{8}{16}  \\ \hline
\end{tabular}
\caption{Results for the beta functions $a\partial
  c_i/\partial a$ and mass derivatives $M\partial c_i/\partial M$,
  computed from fits to the isotropic data sets.}
\label{tab:beta_func}.
\end{table}

The results of the fits to \eqref{eq:xifit}--\eqref{eq:ximfit} are
shown in Table~\ref{tab:anisofits}.  We see that the $\chi^2$ per
degree of freedom is very high, especially for the average
anisotropy and the mass ratio fits.  This indicates that our linear
approximation breaks down in this region, something which in the case
of the anisotropy may be seen directly from the numbers in
Table~\ref{tab:aniso-params}, where a nonlinear response of the
physical anisotropies (and, indeed the lattice spacing) to the bare
anisotropies is evident.  To account for this, we would need to either
include nonlinear terms in our {\em Ansatz} or employ smaller
anisotropies (which would again require much higher statistics to
determine the coefficients with sufficient precision).  That is beyond
the scope of this study.

The generalised Karsch coefficients are presented in
Table~\ref{tab:karsch}.  We see that although the anisotropy
derivatives are reasonably well determined, other quantities,
including the beta functions, have quite large uncertainties.  The
same has been found previously in real QCD with anisotropic lattices
\cite{Morrin:2009}.  It is likely that the extraction of the lattice
spacing from the static quark potential is the main limiting factor
here, and that a high-precision lattice spacing determination from for
example the Wilson flow \cite{Borsanyi:2012zs} (which may also be used
to determine the gauge anisotropy \cite{Borsanyi:2012zr}) would help in this
respect.

A surprising result is the small value for the coefficient
$\partial\gamma_q/\partial\xi$, which comes out between 0.1 and 0.2,
in contrast to $\partial\gamma_g/\partial\xi$, which has a value close
to 1 as expected.  It is possible that this is related to the
breakdown of the linear approximation, and that including non-linear
terms might bring this coefficient closer to 1.  As we shall see in
Sec.~\ref{sec:thermo}, this has a significant impact on the resulting
energy density.

The coefficients $a\partial\gamma_{g,q}/\partial a$ should be zero in
the isotropic limit.  While consistent or nearly consistent with zero
within errors, the central values in Table~\ref{tab:karsch} are fairly
large.  If we could constrain these to be exactly zero, our overall
uncertainties might be reduced.  We also see that
Eq.~\eqref{eq:karsch-beta} is satisfied within the admittedly large
uncertainties.  Again, it might improve the accuracy of our
determination if these equations could be constrained to hold
exactly.

We may also use $m_\pi/m_\rho$ instead of $M=(m_\pi/m_\rho)^2$ as our
mass observables in the fits.  We find that repeating the analysis
above with this choice does not change the results for the Karsch
coefficients and beta functions by much.

We have also computed the beta
functions separately from a 2-dimensional fit to the isotropic
ensembles only.  The results are shown in Table~\ref{tab:beta_func}.
As we can see, the two approaches give consistent results, suggesting
that the systematic uncertainties of the method are under reasonable
control.  The numbers are also roughly consistent with (but somewhat
larger than) the crude estimates used in Ref.~\cquarkyonic, where a
simple backward derivative approximation was used.

\section{Results at $\mu\neq0$}
\label{sec:results}

We now focus on the $(\beta=1.9,\kappa=0.1680)$ parameter set, and
explore the interior of the $(T,\mu)$ plane for these bare couplings.
Results for $j=0.04$
on the $12^3\times24$ lattices were already presented in \cquarkyonic.
Now, with the addition of data for $j=0.02$ and, for some selected
$\mu$-values, $j=0.03$, we can extrapolate all our results to the
$j=0$ limit. The details of this extrapolation will be
discussed in Section~\ref{sec:technical}, as will our treatment of
finite lattice spacing and finite volume lattice artefacts.

\begin{figure}[tb]
\includegraphics*[width=\colw]{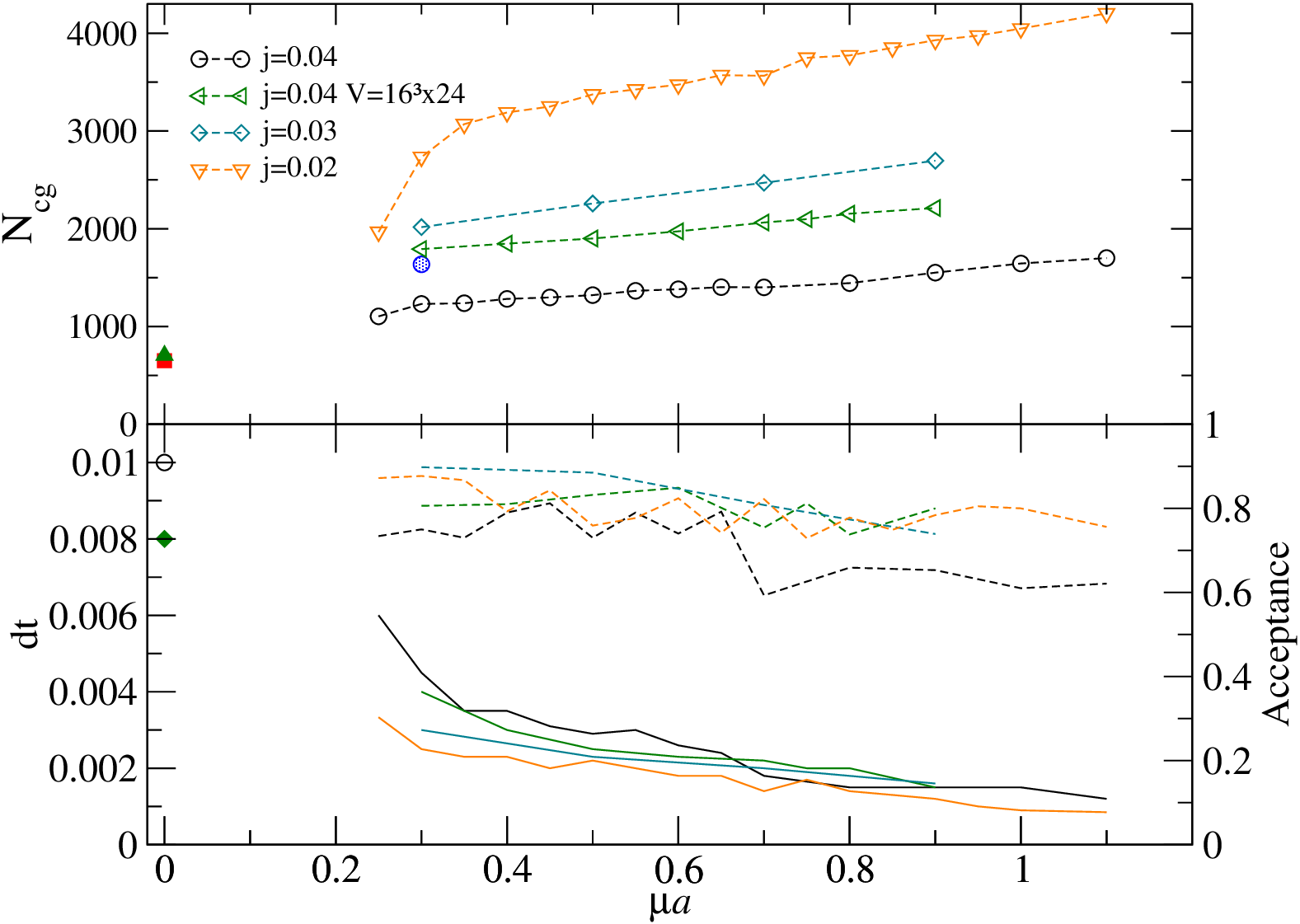}
\caption{The number of conjugate gradient iterations $N_{cg}$ per
  inversion, step size $dt$ (solid lines) and acceptance rates (dashed
  lines) for our simulations on $N_\tau=24$ lattices.}
\label{fig:effort}
\end{figure}

We have also explored higher temperatures with data at
$N_\tau=16,12,8$, and studied finite volume effects with the addition of a
$16^3$ spatial volume.  The temperatures are $T=47$, 70, 94 and 141 MeV for
$N_\tau=24$, 16, 12 and 8 respectively.  Details of our data sets
are given in Tables~\ref{tab:params-mu-t24}--\ref{tab:params-mu-t8}.  
Figure~\ref{fig:effort} shows the computational effort for the
$N_\tau=24$ lattices in terms of the number of conjugate gradient
iterations per inversion and the molecular dynamics stepsize.  It is
evident from this figure that simulations in the dense region at the
lowest $j$-value are 1--2 orders of magnitude more costly than those of
the vacuum.

\begin{table}
\begin{tabular*}{\colw}{l@{\extracolsep{\fill}}rrrr}
\hline
$a\mu$ & $aj=0.02$ & $aj=0.03$ & \multicolumn{2}{c}{$aj=0.04$}
\\ \cline{4-5} 
 & & & $N_s=12$ & $N_s=16$ \\ \hline
 0.25  & 250 &     & 560 & \\
 0.30  & 514 & 315 & 632 & 500 \\
 0.325 & 250 &     & 560 & \\
 0.35  & 284 &    & 1248 & \\
 0.375 & 250 &     & 660 & \\
 0.38  &     &     & 552 & \\
 0.40  & 256 &     & 712 & 500 \\
 0.425 & 264 &     & 592 & \\
 0.45  & 368 &     & 768 & \\
 0.46  &     && 680 \\
 0.47  &     && 468 \\
 0.48  &     && 712 \\
 0.49  &     && 716 \\
 0.50  & 253 & 270 & 699 & 730 \\
 0.525 &     &     & 556 & \\
 0.55  & 260 &     & 168 & \\
 0.575 &     && 314 & \\
 0.60  & 256 && 172 & 510 \\
 0.65  & 260 && 644 &  \\
 0.70  & 253 & 250 & 476 & 560 \\
 0.75  & 255 &     &     & 600 \\
 0.80  & 257 &     & 616 & 600 \\
 0.85  & 255 &&& \\
 0.90  & 250 & 260 & 316 & 560 \\
 0.95  & 257 &&& \\
 1.00  & 250 && 600 \\
 1.10  & 252 && 504 \\\hline
\end{tabular*}
\caption{Number of trajectories for $\mu\neq0$, $\beta=1.9,
  \kappa=0.168, N_\tau=24$ ($T=47$ MeV).  The $ja=0.02, 0.03$
  configurations all have $N_s=12$.  All trajectories have average
  length 0.5.} 
\label{tab:params-mu-t24}
\end{table}

\begin{table}
\begin{tabular}{l|rrrrrrrrr}
\hline
$a\mu$ & 0.300 & 0.400 & 0.450 & 0.500 & 0.525 & 0.550 & 0.575 \\
$\ntr$ & 500   &   560 &  2000 &  2520 &  2045 &  2000 & 2550 \\
\hline
$a\mu$ & 0.600 & 0.625 & 0.650 & 0.675 & 0.700 & 0.800 & 0.900 \\
$\ntr$ &  2520 &  2520 &   560 &   560 &   520 &   540 &   500 \\
\hline
\end{tabular}
\caption{Chemical potential values and number of trajectories for the
  $12^3\times16$ lattices ($T=70$ MeV).  The diquark source is $ja=0.04$ in all
  cases.  All trajectories have average length 0.5.}
\label{tab:params-mu-t16}
\end{table}

\begin{table}
\begin{tabular}{l|rrrrrrrrrrr}
\hline
$a\mu$ & 0.200 & 0.250 & 0.275 & 0.300 & 0.325 & 0.350 & 0.360 & 0.375 & 0.390 \\
$N (0.04)$ & 1000  &  2500 &  2520 &  2520 &  2800 &  4900 &  2100 &  4900 &  1300 \\
\hline
$a\mu$ & 0.400 & 0.425 & 0.450 & 0.500 & 0.600 & 0.700 & 0.800 & 0.900 \\
$N(0.04)$ &  1080 &  1050 &  1050 &  1164 &  1128 & 600 & 510 & 540 \\
$N(0.02)$ & 500 &  &  & 512 & 310 & 300 & 250 & 255 \\
\hline
\end{tabular}
\caption{Chemical potential values and number of trajectories for the
  $16^3\times12$ lattices ($T=94$ MeV). $N(0.04)$ and $N(0.02)$ are
  the number of trajectories for $ja=0.04$ and 0.02 respectively. All
  trajectories have average length 0.5.}
\label{tab:params-mu-t12}
\end{table}

\begin{table}
\begin{tabular}{l|rrrrrrrrrrr}
\hline
$a\mu$ & 0.100 & 0.200 & 0.300 & 0.400 & 0.500 & 0.600 & 0.700 & 0.800 & 0.900\\
$N(0.04)$ & 1000 & 1000 & 1000 & 1050 & 1050 & 1200 & 1000 & 1000 & 1000 \\
$N(0.02)$ &  &  &  & 1000 & 1000 & 1000 & 1000 & & \\
\hline
\end{tabular}
\caption{As Table~\ref{tab:params-mu-t12}, for the $16^3\times8$
  lattices ($T=141$ MeV).}
\label{tab:params-mu-t8}
\end{table}

\subsection{Diquark source extrapolation and lattice artefacts}
\label{sec:technical}

\begin{figure}
\includegraphics*[width=\colw]{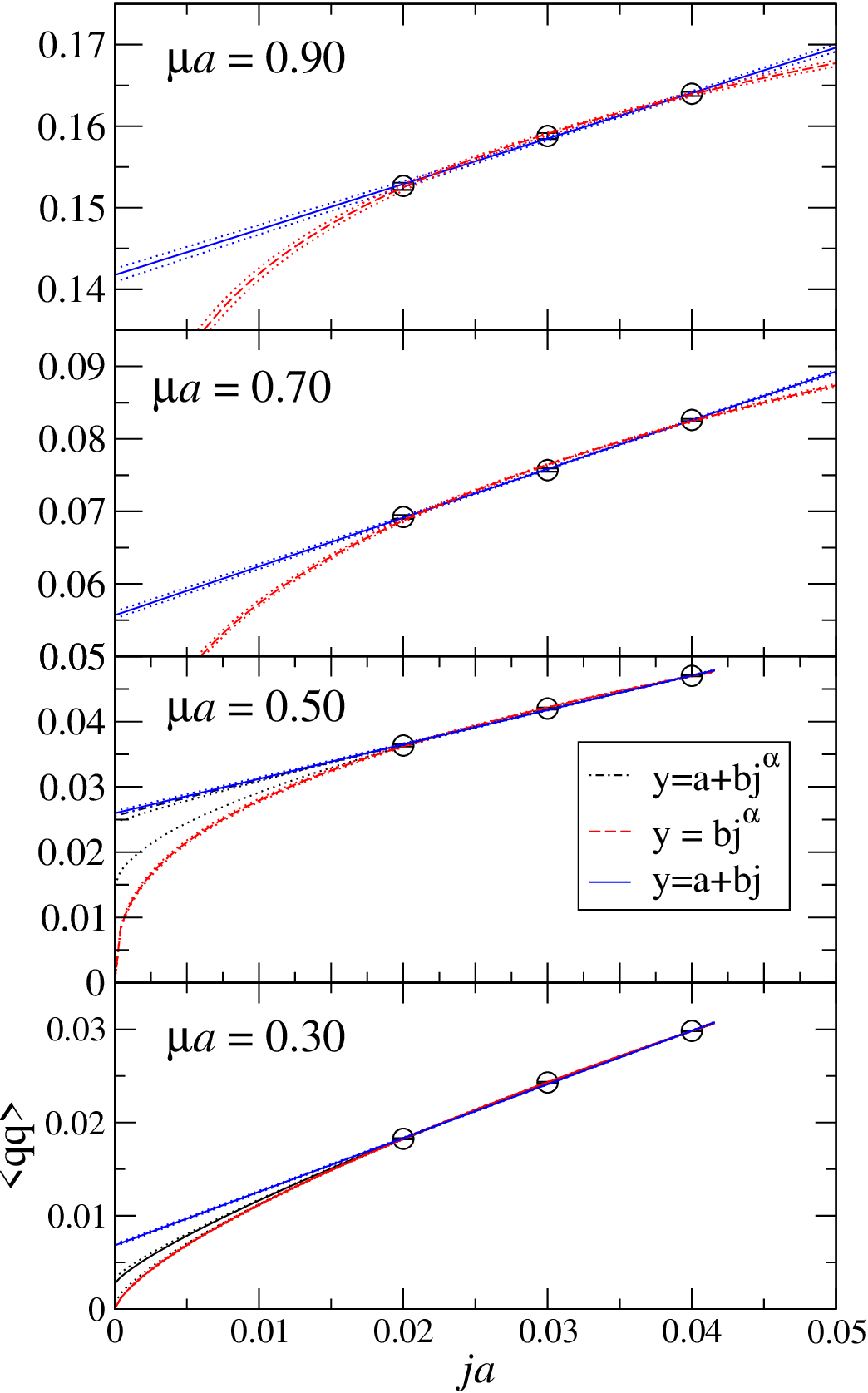}
\caption{The diquark condensate $\qq$ as a function of diquark source
  $j$, for the $12^3\times24$ lattice, together with extrapolations to
  $j=0$.  The dotted lines denote the
  68\% confidence interval for each fit.  At $\mu=0.50$ the central
  value lies outside the 68\% confidence interval.}
\label{fig:diquark-extrapolate}  
\end{figure}

In Fig.~\ref{fig:diquark-extrapolate} we show the diquark
condensate $\qq$ as a function of the diquark source $j$ for 
$\mu a=0.3, 0.5, 0.6, 0.9$ on the $12^3\times24$ lattice.  We have
attempted to fit the behaviour 
with three different functional forms: linear ($\qq=A+Bj$), power-law
($\qq=Bj^\alpha$) and constant + power ($\qq=A+Bj^\alpha$).  Our
results are summarised in Table~\ref{tab:extrap-qq}.  We find
that a linear fit works reasonably well except for $\mu a=0.3$, where
a pure power-law works well, confirming that the diquark condensate is
indeed zero at this point.  At $\mu a=0.5$, neither functional form
gives a very good fit, but the constant + power fit gives a result for
the extrapolated diquark condensate consistent with the linear form.
Note that the constant + power fit is always far less stable than the
two others, but for $\mu a\geq0.5$ the extrapolated values are
consistent with those from the linear fit.

\begin{table}
\begin{center}
\begin{tabular}{|l|cccc|}
\hline
$\mu a$ & 0.3 & 0.5 & 0.7 & 0.9 \\\hline\hline
\multicolumn{5}{|c|}{Linear fit $A+Bj$} \\\hline
$A$ & 0.0068(1) & 0.0260(3) & 0.0557(5) & 0.1418(8) \\
$\chi^2$ & 7.5 & 3.3 & 0.06 & 1.05 \\ \hline
\multicolumn{5}{|c|}{Power law fit $Bj^\alpha$} \\\hline
$\alpha$ & 0.709(6) & 0.376(7) & 0.261(6) & 0.104(5) \\
$\chi^2$ & 0.23 & 2.1 & 15.1 & 1.03 \\\hline
\multicolumn{5}{|c|}{Power + constant fit $A+Bj^\alpha$} \\\hline
$A$ & 0.0027\err{5}{21} & 0.025\err{-1}{11} & 0.058\err{2}{3} 
 & 0.129\err{6}{13} \\
$\alpha$ & 0.36\err{2}{6} & 0.50\err{-7}{29} & 1.0\err{1.0}{0.4}
 & 0.21\err{7}{4} \\ \hline
\end{tabular}
\end{center}
\caption{Parameters for $j\to0$ extrapolations of the diquark
  condensate $\qq$.  Note that the power + constant fit is a 3-parameter fit
to 3 data points, and hence there is no $\chi^2$ for this fit.}
\label{tab:extrap-qq}
\end{table}

\begin{table}
\begin{center}
\begin{tabular}{|l|cccc|}
\hline
$\mu a$ & 0.3 & 0.5 & 0.7 & 0.9 \\\hline\hline
\multicolumn{5}{|c|}{Linear fit $A+Bj$} \\\hline
$A$ & 0.0000(5) & 0.0128(9) & 0.0407(17) & 0.190(3) \\
$\chi^2$ & 6.8 & 3.7 & 0.04 & 0.04 \\ \hline
\multicolumn{5}{|c|}{Power law fit $Bj^\alpha$} \\\hline
$\alpha$ & 0.95(17) & 0.18(5) & 0.20(3) & 0.076(15) \\
$\chi^2$ & 6.7 & 4.8 & 0.07 & 0.11 \\\hline
\multicolumn{5}{|c|}{Power + constant fit $A+Bj^\alpha$} \\\hline
$A$ & 0.000\err{-5}{24} & 0.0162\err{2}{2} & 0.025\err{11}{46} 
 & 0.185\err{9}{64} \\
$\alpha$ & 0.21\err{-18}{19} & -0.0171\err{-13}{0} & 0.11\err{8}{1}
 & 0.29\err{47}{13} \\ \hline
\end{tabular}
\end{center}
\caption{Parameters for $j\to0$ extrapolations of the quark number
  density $n_q$, for the $12^3\times24$ lattice.}
\label{tab:extrap-nq}
\end{table}

The results for other observables are similar.  As an illustration of
this, the corresponding fits for the quark number density $n_q$ summarised
in Table~\ref{tab:extrap-nq}.  Based on these findings, we use a
linear function as our default extrapolation model for all
observables, keeping in mind that this will distort the results
somewhat in the r\'egime $\mu a\lesssim0.5$.

Next, we discuss our treatment of lattice artefacts in the context of
the quark number density $n_q$.
As in previous works, it will prove convenient to express results in
terms of dimensionless ratios, eg. $n_q/n_q^{SB}$, where $n_q^{SB}$ is
the result for non-interacting quarks. However, even for free quarks
artifacts due to non-zero lattice spacing and finite spatial volume
are non-negligible, resulting in significant departures from the
result in continuum and thermodynamic limits, and very careful
discussion is required.
\begin{figure}[htb]
\includegraphics*[width=\colw]{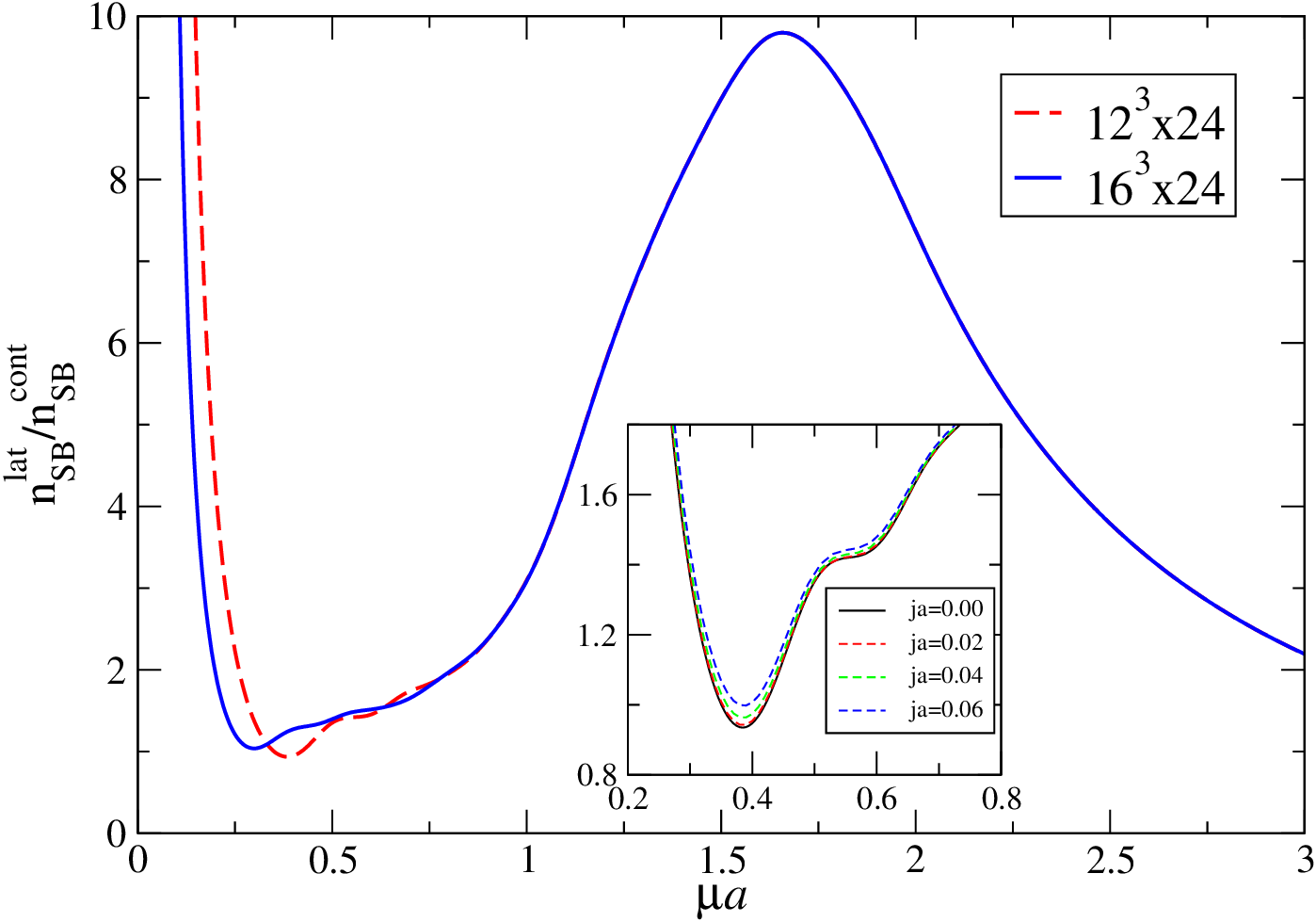}
\caption{Ratio $n_{SB}^{\rm lat}/n_{SB}^{\rm cont}$ evaluated for free
  massless quarks on both $12^3\times24$ and $16^3\times24$ lattices.
  The inset shows the same ratio for the $12^3\times24$ lattice, for
  four different values of the diquark source $j$.}
\label{fig:freefermion}
\end{figure}
Insight into both UV and IR artefacts can be
gleaned by considering the ratio
$n_{SB}^{\text{lat}}/n_{SB}^{\text{cont}}(T=0)$, calculated  for two
different volumes using the formula
given in \cdeconf, and shown in Fig.~\ref{fig:freefermion}. 
The correction is numerically large across extensive portions
of the $\mu$-axis. The oscillatory behaviour seen for $\mu a<0.8$ is
an IR artefact known to arise from the non-sphericity of the Fermi
surface  resulting from the discretisation of momentum
space~\cite{Hands:2002mr}.

\begin{figure}[tb]
\includegraphics*[width=\colw]{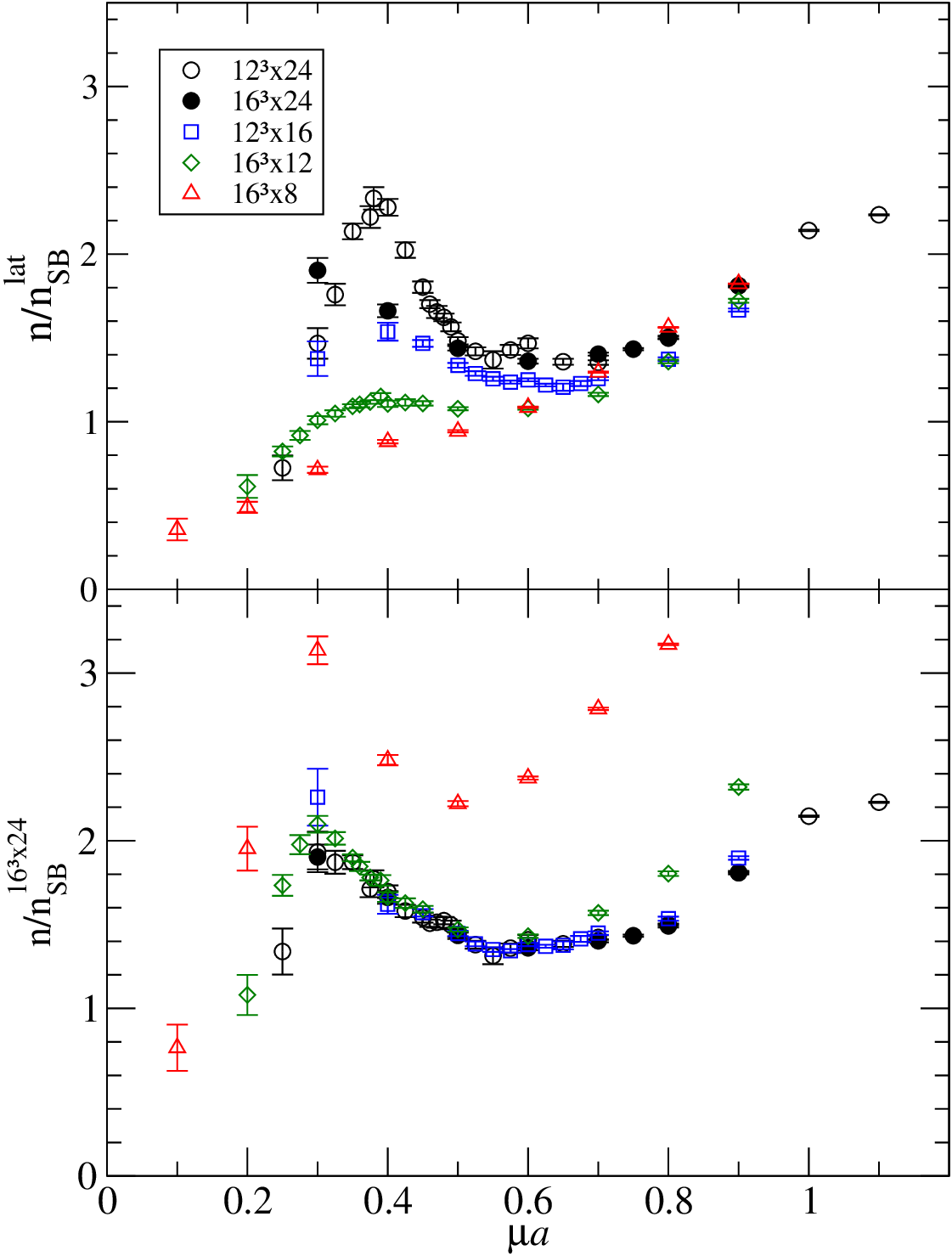}
\caption{The quark number density at $ja=0.04$ for different lattice
  volumes, divided by the density for a noninteracting gas of lattice
  quarks on the same volume (top) and on a fixed volume of
  $16^3\times24$ (bottom).}
\label{fig:density-compare}
\end{figure}
As an illustration of these effects, in Fig.~\ref{fig:density-compare}
we show the normalised quark number density $n_q/n_{SB}$ at fixed
diquark source $ja=0.04$, with two different choices for $n_{SB}$.
In the upper panel we have
normalised by $n_{SB}$ for the corresponding lattice volumes, while in
the lower panel we have used the same normalisation for all lattices.
We have chosen to use $n_{SB}$ for a $16^3\times24$ lattice for this
normalisation; note that this choice is purely a matter of
convenience, the purpose being to easily compare the raw numbers for
$n_q$ from different lattices.  We see that there is no difference
between our raw numbers for $n_q$ on the $12^3\times24$ and
$16^3\times24$ lattices at $j=0.04$; however $n_{SB}$ for the $12^3$
lattice has a dip around $\mu a\simeq0.4$, while on the
$16^3$ lattice this feature has moved to smaller $\mu$.  This dip
coincides with the peak in $n_q/n_{SB}$ seen in the 
upper panel of Fig.~\ref{fig:density-compare}, giving rise to a
spurious discrepancy in the normalised results for the two volumes.

By contrast, the correction factor coincides
on the two volumes for $\mu a > \order(1)$, suggesting that the
considerable departure from unity at large $\mu$  is due to UV
effects.  As we can see in the inset of Fig.~\ref{fig:freefermion},
the diquark source has a negligible effect on the noninteracting quark
density, and hence any significant $j$-dependence in our results
must arise from interactions.

Based on these findings, we will in the following present our results
for $n_q$ and the pressure $p$, as well as the quark number
susceptibility $\chi_q$, using both the noninteracting lattice and
continuum expressions to normalise our data.  This will allow us to
assess the magnitude of IR and UV lattice artefacts.  For the energy
density and trace anomaly, where gluonic contributions are
significant, we will instead normalise by $\mu^4$.

\subsection{Order parameters and phase structure}
\label{sec:phases}

\begin{figure}[tb]
\includegraphics*[width=\colw]{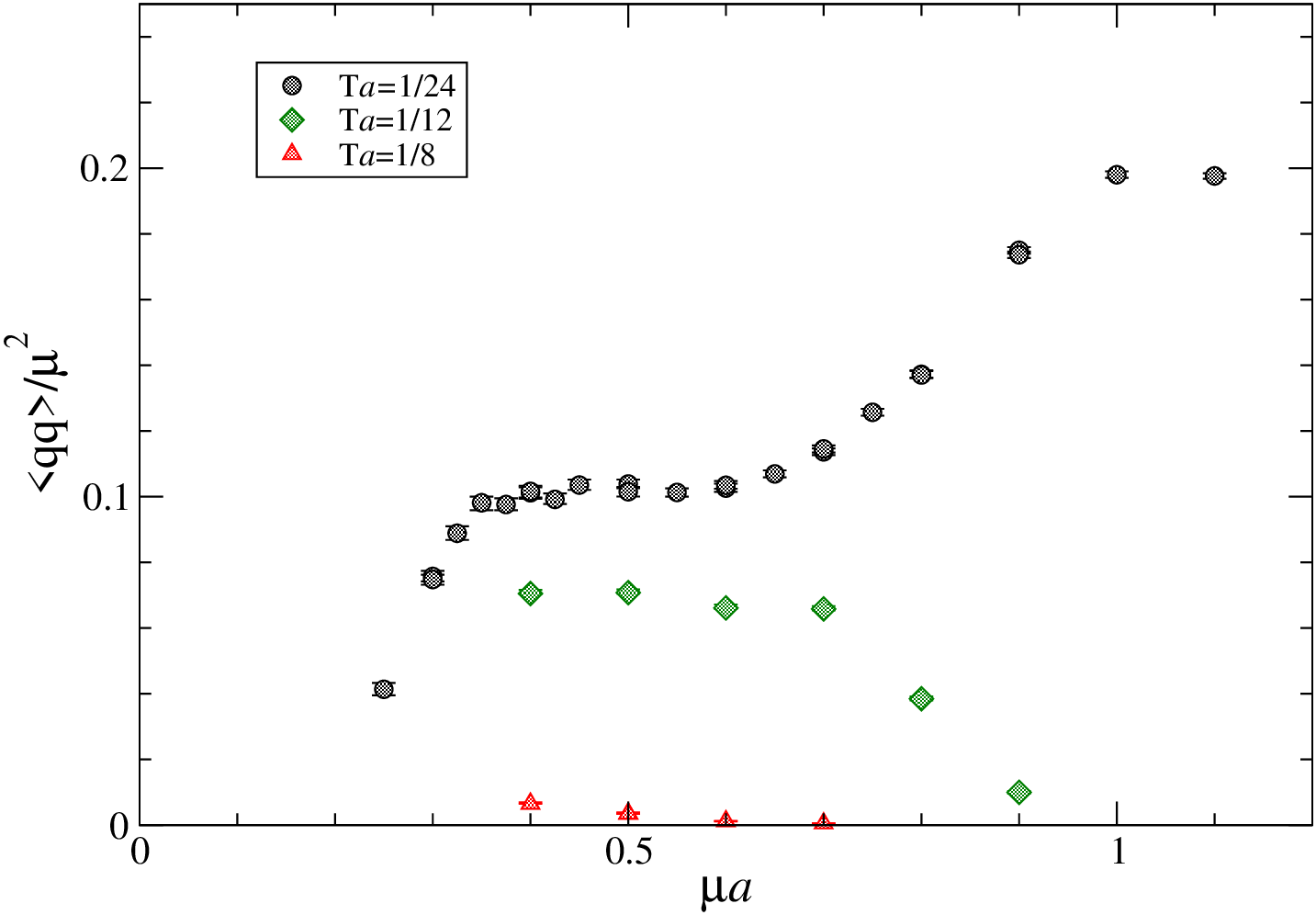}
\caption{The diquark condensate $\bra qq\ket/\mu^2$
  extrapolated to $j=0$ for $N_\tau=24, 12, 8$ ($T=47,94,141$ MeV).}
\label{fig:qq}
\end{figure}

Figure~\ref{fig:qq} shows the diquark condensate,
\begin{equation}
\qq = \bra\psi^{2tr}C\gamma_5\tau_2\psi^1
  -\bar\psi^1C\gamma_5\tau_2\bar\psi^{2tr}\ket\,,
\label{def:qq}
\end{equation}
as a function of chemical potential, for the $N_\tau=24, 12$ and 8
lattices.  In the case of a weakly-coupled BCS condensate at the Fermi
surface, the diquark condensate, which is the number density of Cooper
pairs, should be proportional to the area of the Fermi surface, ie
$\qq\sim\mu^2$. This is to be contrasted with chiral perturbation
theory ($\chi$PT) \cite{Kogut:2000ek}, which for $\mu\gg\mu_o$ at
leading order predicts $\bra qq\ket$ to be $\mu$-independent.

For the lowest temperature $T=47$ MeV ($N_\tau=24$) we see an almost perfect
proportionality in the region $0.35\lesssim\mu a\lesssim0.6$.  The
lower limit of this region roughly coincides with the onset chemical
potential $\mu_o\approx m_\pi/2\approx0.33a^{-1}$, below which both
the quark number density and diquark condensate is expected to be
zero.  The reason we see a gradual rise from $\mu a\approx0.25$ is our
use of a linear Ansatz for the $j\to0$ extrapolation, which is not
valid in this r\'egime, as discussed in Section~\ref{sec:technical}.
For $\mu a\gtrsim0.6$, $\qq/\mu^2$ rises again before possibly
reaching a new plateau at $\mu a\approx1.0$.  This is possible evidence
of a transition to a new state of matter at high density, but at these
large densities the impact of lattice artifacts cannot be excluded.

At $T=70$ MeV ($N_\tau=16$) we are not in a position to perform a $j\to0$
extrapolation, but from the $ja=0.04$ data we see only a mild
suppression in $\qq$, and only for $\mu a\gtrsim0.8$. Since
the results are almost indistinguishable from those at $T=47$ MeV we
do not show them here.

At $T=94$ MeV ($N_\tau=12$) we see that $\qq$ is significantly smaller
for all values of $\mu$ and drops dramatically above $\mu
a\gtrsim0.7$.  This gives us the first indications of the transition
between the diquark-condensed and the normal phase.  At $T=141$ MeV
($N_\tau=8$) we
find that the diquark condensate is zero at all $\mu$, confirming that
the system is in the normal phase at this temperature.  A systematic
investigation including more temperatures and an extrapolation to
$j=0$ at all temperatures will be required to establish the exact
location and nature of this transition.

Finally, comparing the numbers from the $12^3\times24$ and
$16^3\times24$ lattices, no evidence of any significant finite volume
effects are found, except at $\mu a=0.9$ where the condensate on the
smaller volume is slightly suppressed.

\begin{figure}[tb]
\includegraphics*[width=\colw]{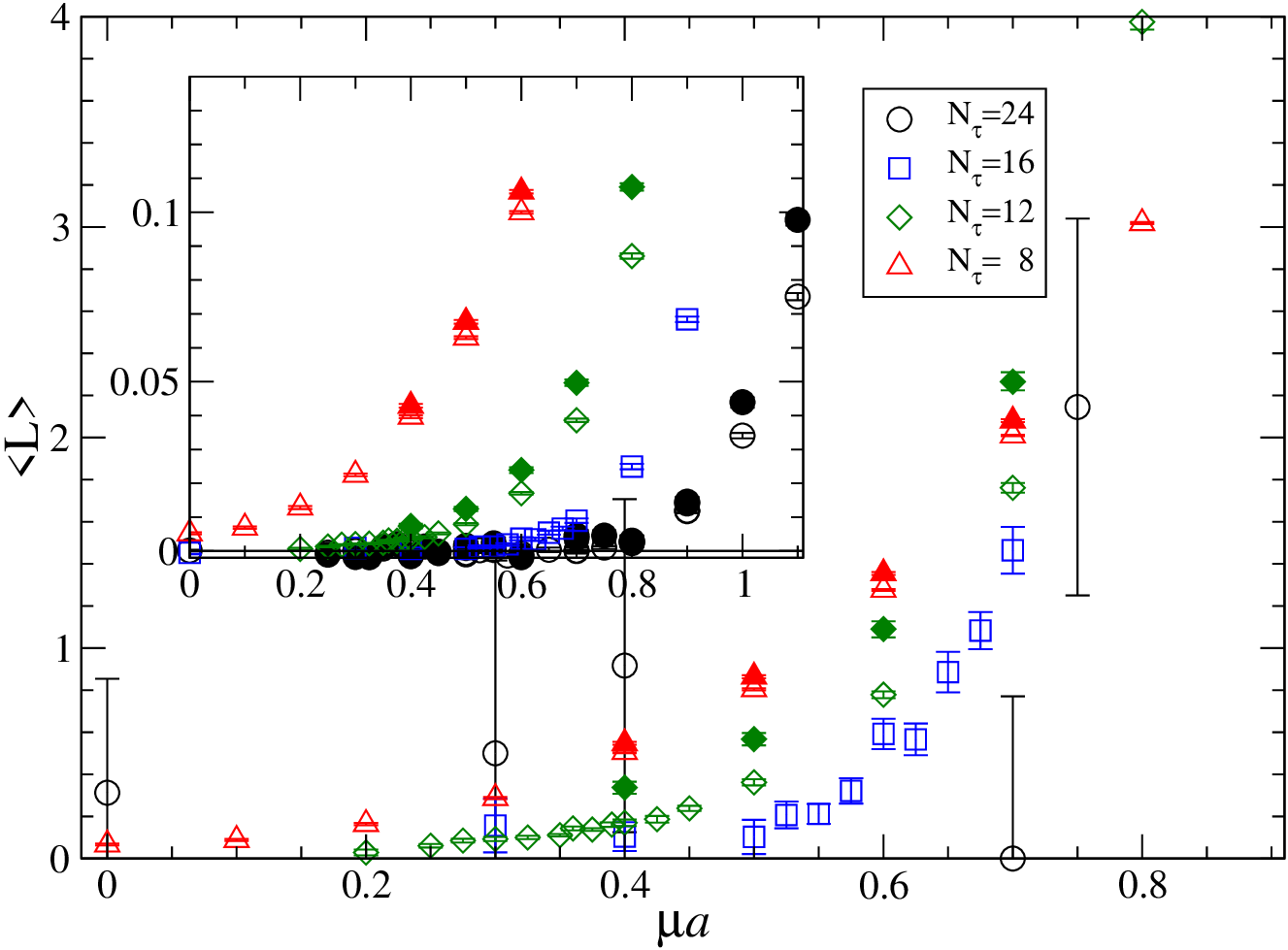}
\caption{The renormalised Polyakov loop as a function of chemical
  potential, for all temperatures.  The open symbols are for
    $ja=0.04$; the filled symbols are extrapolated to $j=0$.  The
  inset shows the unrenormalised Polyakov loop.}
\label{fig:polyakov}
\end{figure}

\begin{figure}[tb]
\includegraphics*[width=\colw]{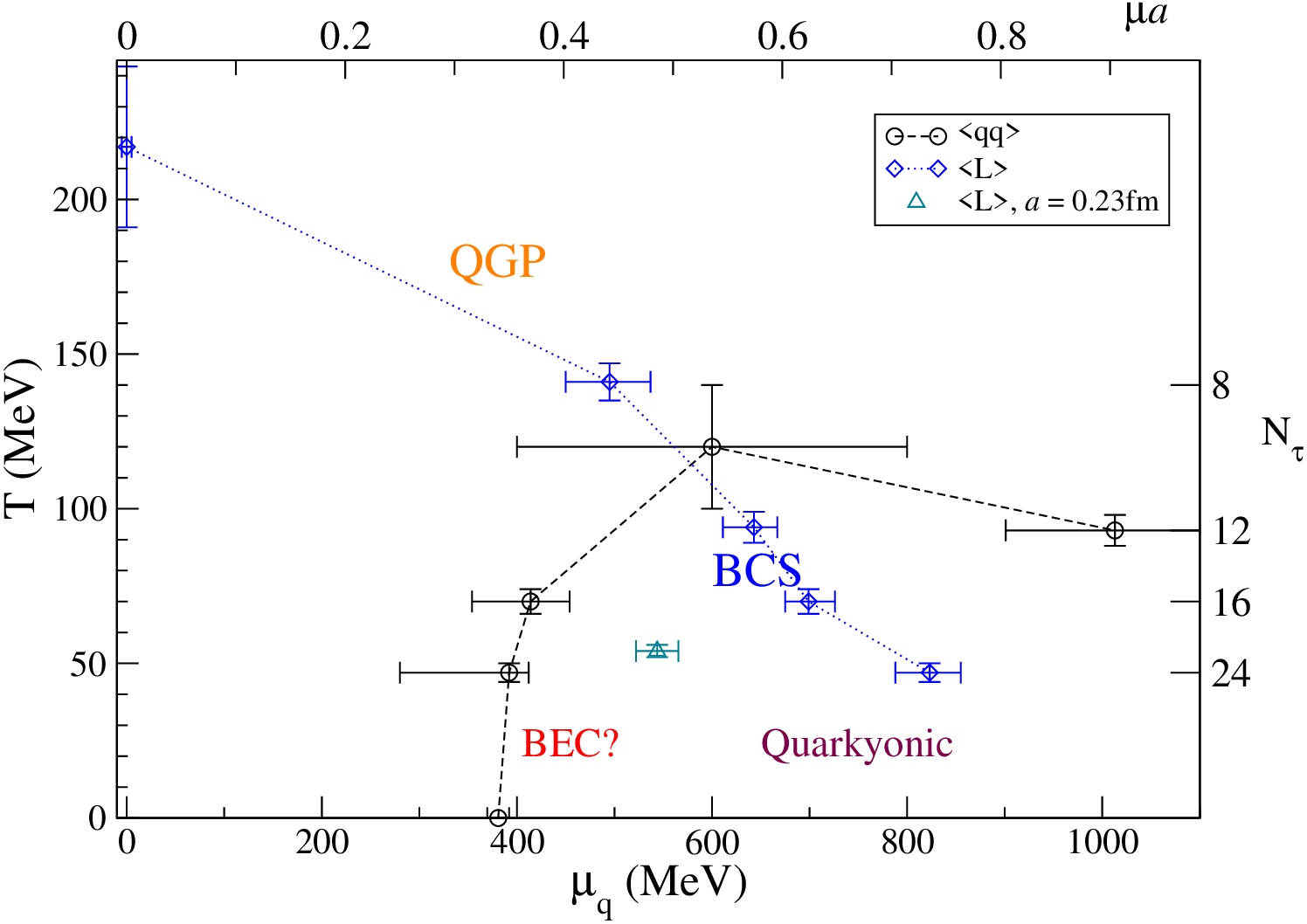}
\caption{A tentative phase diagram, including the location of the
  deconfinement transition in the $(\mu,T)$ plane, determined from the
  renormalised Polyakov loop, and the transition to the diquark
  condensed $\bra qq\ket\neq0$ phase.  Also shown is the deconfinement
  point from Ref.~\cdeconf.}
\label{fig:phases}
\end{figure}

Figure~\ref{fig:polyakov} shows the order parameter for deconfinement,
the Polyakov loop $\bra L\ket$, for our four different temperatures.
It has been renormalised using \eqref{eq:polyakov-renorm},
using the $\mu$-independent renormalisation constant $Z_L$ already
computed in Sec.~\ref{sec:vacuum}.  We see that for each
temperature $T$, $\bra L\ket$ increases rapidly from zero above a
chemical potential $\mu_d(T)$ which we may identify with the chemical
potential for deconfinement.  However, since $L$ is a convex function
of $\mu$ at all $T$, it is not possible to use the variation of $L$
with $\mu$ to define $\mu_d(T)$.  In the absence of a more rigorous
criterion, we have taken the point where $L$ crosses the value it
takes at $T_d(\mu=0)$, $L_d=0.6$, to define $\mu_d(T)$.  The results
are shown in Fig.~\ref{fig:phases}, with error bars denoting the range
$L_d=$0.5--0.7.  To more accurately locate the deconfinement line, we
will need to perform a temperature scan for fixed $\mu$-values, as was
done for $\mu=0$.

For our lowest temperature ($N_\tau=24)$, the renormalised Polyakov
loop is too noisy for any quantitative conclusions to be drawn.  This
is because the signal (which is consistent with 0 for $\mu a<0.75$) as
well as the statistical noise are multiplied by the large
renormalisation factor $Z_L^{24}=2084$.  However, the unrenormalised
Polyakov loop $L_0$, shown in the inset of Fig.~\ref{fig:polyakov},
exhibits the same qualitative behaviour as for the higher
temperatures.  We also find that there are no significant volume
effects, while the diquark source tends to suppress the Polyakov loop
slightly.  At $\mu a\approx0.75$ we see that the curves for the
renormalised Polyakov loop at the different temperatures cross, so
that at higher $\mu$, $L$ is smaller for higher temperatures.  This,
however, depends on the renormalisation scheme: if we had instead
imposed the condition that $L_R=0.5$ at $N_\tau=4,\mu=0$, the curves
would not cross.

The estimates of critical chemical potentials for both deconfinement
and superfluidity can be translated into a tentative phase diagram,
shown in Fig.~\ref{fig:phases}.  It is worth reiterating that the
points on the phase boundaries are rough estimates only, since we do
not have a precise criterion for the transition.  In
Section~\ref{sec:susc} we will present results for a different measure
of deconfinement, the quark number susceptibility.  We also show the
estimate from the coarser lattice in Ref.~\cdeconf.  Clearly, a
combination of temperature effects and lattice artefacts is
responsible for the discrepancy between the $\mu_d$-values quoted in
\cprev.

In Fig.~\ref{fig:phases} we also show our estimate of the transition
between the superfluid and the normal phase.  Again, since we do not
yet have $j\to0$ extrapolated data at all temperatures, and because
our temperature grid is fairly coarse, these transition points are
also only rough estimates.

In summary, from the order parameters we find signatures of three
different regions (or phases): a normal (hadronic) phase with
$\qq=0,\braket{L}\approx0$; a BCS (quarkyonic) region with
$\qq\sim\mu^2$ at low $T$ and intermediate to large $\mu$; and a
deconfined, normal phase with $\qq=0,\braket{L}\neq0$ at large $T$
and/or $\mu$. We cannot exclude a deconfined superfluid phase with
$\bra L\ket>0$, $\bra qq\ket\not=0$ at large $\mu$ and intermediate
$T$.

After extrapolating our results to zero diquark source, we see no
evidence of a BEC region described by $\chi$PT, with
$\qq\sim\sqrt{1-\mu_o^4/\mu^4}$ \cite{Kogut:2000ek}, in contrast with
earlier work with staggered lattice fermions \cite{Hands:2000ei}.
This may be because we do not have a clear separation of scales
between the Goldstone diquark scale and more massive states, and hence
the region of tightly bound diquarks is very narrow.  A more
pessimistic scenario is that the BEC region is masked by the poor
chiral properties of Wilson fermions. Simulations with lighter quarks
may help clarify this.

\begin{figure}[tb]
\includegraphics*[width=\colw]{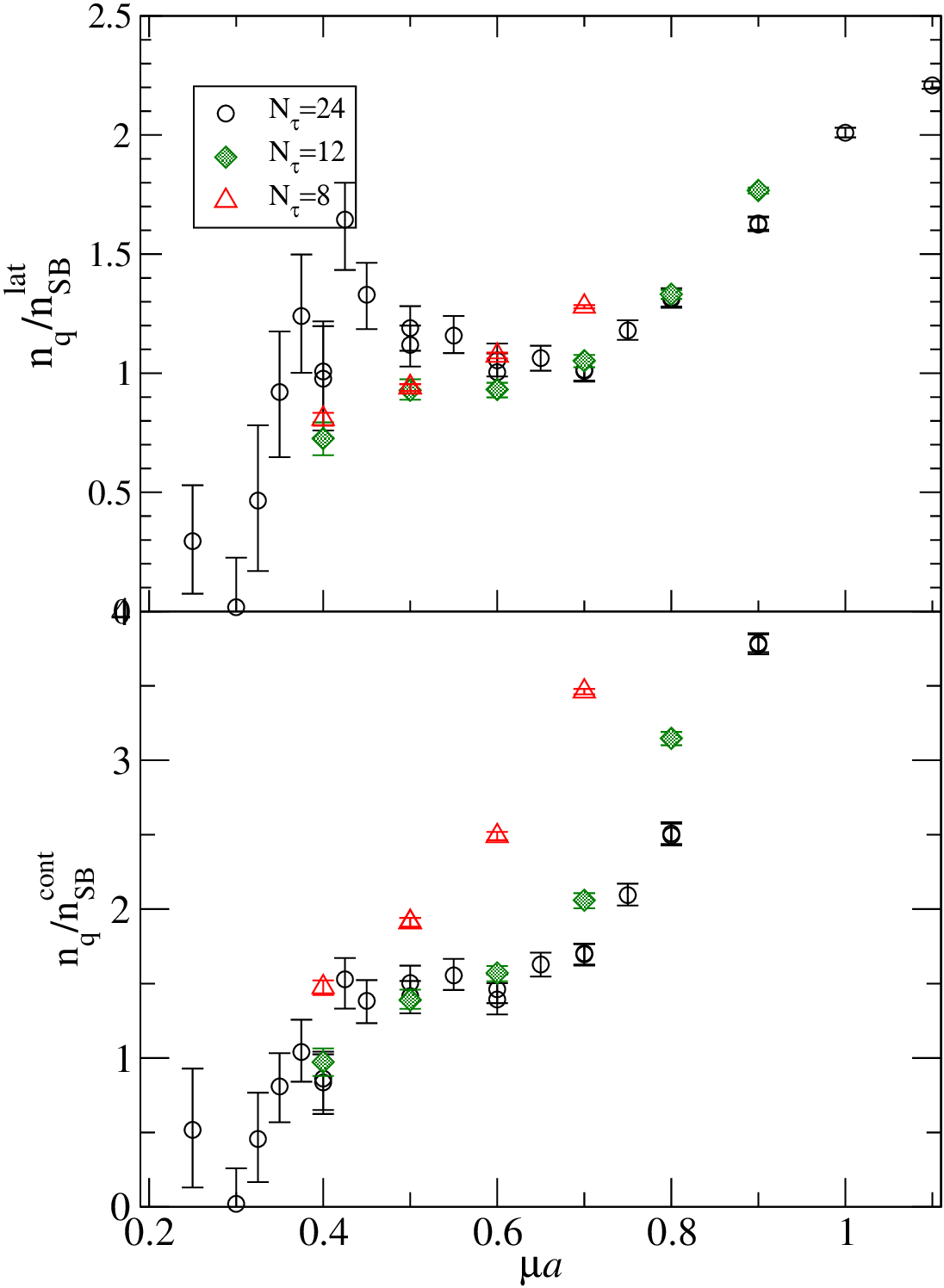}
\caption{The quark number density at $j=0$, divided by the
  density for a noninteracting gas of lattice quarks (top) and
  continuum quarks (bottom).}
\label{fig:density}
\end{figure}

\subsection{Equation of state}
\label{sec:thermo}

We now turn to the bulk thermodynamics of the system: the quark number
density $n_q$, the pressure $p$ and the energy density $\eps$.
Figure~\ref{fig:density} shows the quark number density $n_q$ for
$N_\tau=24, 12$ and 8, extrapolated to zero diquark source. 
In the top panel we have normalised by the density $n_{SB}^{\text{lat}}$ for 
noninteracting fermions on the same lattice volumes ($12^3\times24,
16^3\times12, 16^3$), as was done in
\cprev.  In the bottom panel, we have instead divided by the
continuum, infinite-volume expression for noninteracting fermions at
the same temperature and chemical potential.  The difference between
the two gives an indication of the lattice artefacts.  We see that the 
density rises from zero at $\mu\approx\mu_o=0.32a^{-1}$, and for the
two lower temperatures is roughly constant and approximately equal to
the  noninteracting fermion density in the region $0.4\lesssim\mu
a\lesssim0.7$. The peak  at $\mu a\simeq0.4$ in the $N_\tau=24$ data
in the upper panel is an artefact of the normalisation with $n_{SB}$
for a finite lattice volume, as discussed in
Sec.~\ref{sec:technical}; it would be absent if we instead normalised
by $n_{SB}$ for a $16^3$ lattice, for which the raw data are identical
within errors.  We therefore conclude that our previous
interpretation \cdeconf\ of the peak in $n_q/n_{SB}$ in this region as
evidence of a BEC condensate described by $\chi$PT was probably
erroneous.

Our results for $N_\tau=16$ are indistinguishable from the $N_\tau=24$
results except for $\mu a\gtrsim0.8$, where they also start increasing
above the $N_\tau=24$ values. The rise in $n_q/n_{SB}$ for $\mu
a\gtrsim0.7$ may be a signal of a new phase, although in this region
the influence of lattice artefacts cannot yet be ruled out.

We also note that
$n_q/n_{SB}$ for $N_\tau=12$ rises above the corresponding $N_\tau=24$
data for $\mu a\gtrsim0.7$, where, according to the results of
Sec.~\ref{sec:phases}, the hotter system is entering the deconfined,
normal phase.
The density for $N_\tau=8$ does not show any plateau as a function of
$\mu$; instead, $n_q/n_{SB}$ shows a roughly linear increase in the
region $0.4\leq\mu a\leq0.7$.  This is suggestive of the system being
in a different phase at this temperature.

These results lend further support to our previous conjecture that in
the intermediate-density region the system is in a ``quarkyonic''
phase: a confined phase (all excitations are colourless) that can be
described by quark degrees of freedom.  We reiterate that because of
the large explicit breaking of chiral symmetry in our simulations, we
cannot say anything at this point about chiral symmetry restoration,
another characteristic of the quarkyonic phase conjectured in
Ref.~\cite{McLerran:2007qj}.  We will come back to this issue in
Sec.~\ref{sec:chisb}.

\begin{figure*}[thb]
\includegraphics*[width=0.95\textwidth]{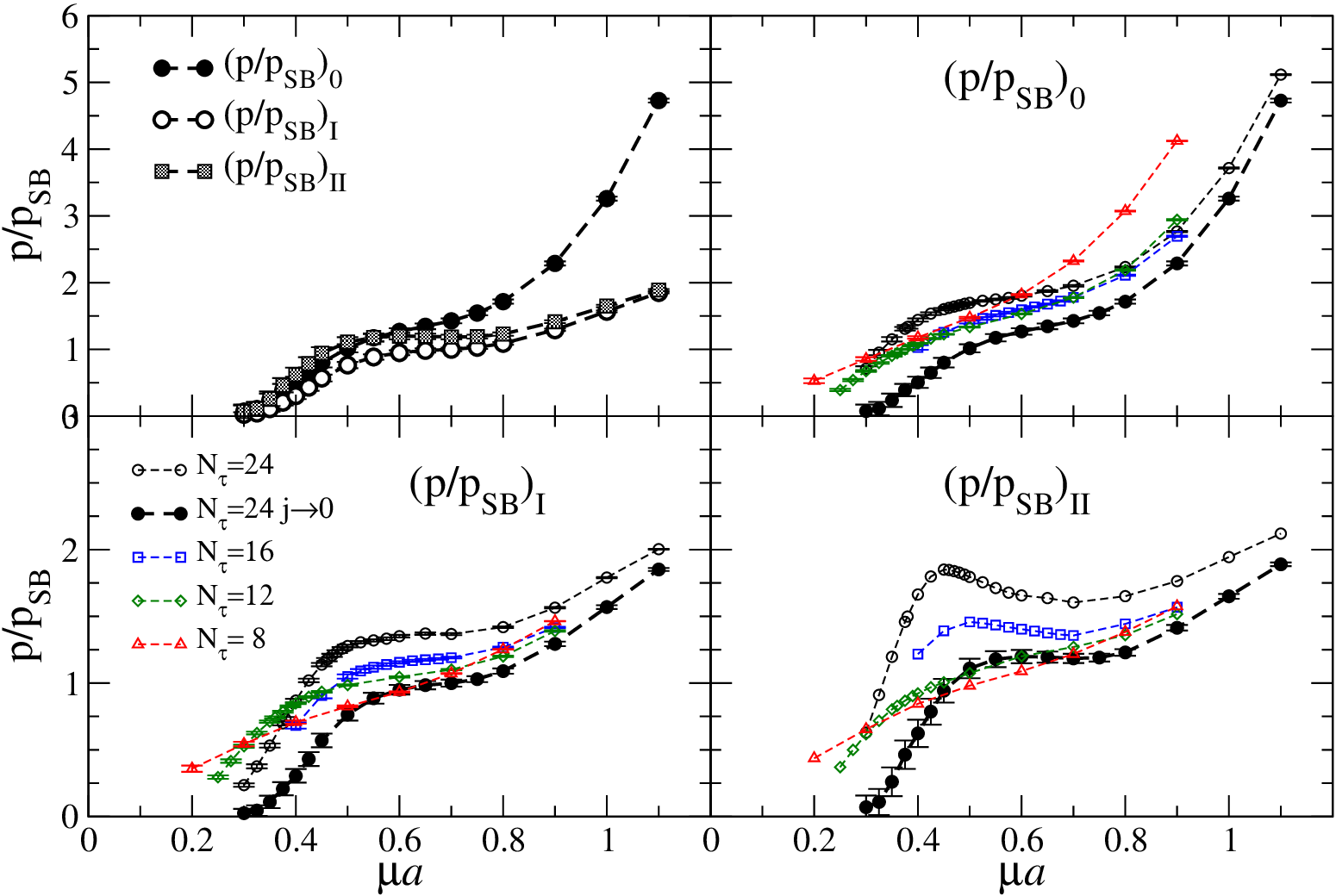}
\caption{$p/p_{SB}$ vs. $\mu a$ for $ja=0.04$ and various
  temperatures.  Also shown are values extrapolated to $j=0$ for
  $N_\tau=24$.  Top left: $(p/p_{SB})_{0,I,II}$
  for $j\to0$, $N_\tau=24$.  Top right: $(p/p_{SB})_0$.  Bottom left:
  $(p/p_{SB})_I$. Bottom right: $(p/p_{SB})_{II}$.}
\label{fig:pressure}
\end{figure*}
Next we discuss pressure, which as the negative of the free energy
density, may be calculated via the integral of any thermodynamic
observable  along an appropriate contour. It is particularly
convenient to integrate along the $\mu$-axis via $p=\int^\mu_{\mu_0}
n_q d\mu$, since the cutoff does not change. Here $\mu_0$ is chosen so
that $p(\mu_0)=0$ to good approximation; in the limit $T\to0$ $\mu_0$
should coincide with the onset $\mu_o$.

In our analysis the integral is readily approximated by a trapezoidal
rule; as always, we present data normalised by the free field value
$p_{SB}$, a procedure  not uniquely defined away from the continuum
limit. We have examined three schemes:
\begin{align}
\left(\frac{p}{p_{SB}}\right)_0
 &=(p_{SB}^{\text{cont}}(\mu))^{-1}\int_{\mu_0}^\mu n_q(\mu^\prime)d\mu^\prime\,,\\
 \left(\frac{p}{p_{SB}}\right)_I
 &=(p_{SB}^{\text{lat}}(\mu))^{-1}\int_{\mu_0}^\mu
   n_q(\mu^\prime)d\mu^\prime\,,\label{eq:pI}\\
 \left(\frac{p}{p_{SB}}\right)_{II}
  &=(p_{SB}^{\text{cont}}(\mu))^{-1}\int_{\mu_0}^\mu
   \frac{n^{\text{cont}}_{SB}}{n^{\text{lat}_{SB}}}(\mu^\prime)
    n_q(\mu^\prime)d\mu^\prime\label{eq:pII}\,,
\end{align}
where
\begin{equation}
p^{\text{cont}}_{SB}=\frac{N_fN_c}{12\pi^2}\left(\mu^4+2\pi^2\mu^2T^2+\frac{7\pi^4}{15}T^4\right)
\end{equation}
is the continuum pressure for a free gas of quarks, and
$p^{\text{lat}}_{SB}$ the corresponding value obtained by summing over
free quark modes on the finite lattice.  Versions (\ref{eq:pI}) and
(\ref{eq:pII}) were both studied in \cdeconf, whereas only
$(p/p_{SB})_{II}$ was used in \cite{Hands:2010gd,Hands:2011ye}.

Fig.~\ref{fig:pressure} shows the results
for data taken with $ja=0.04$, as well as the $j\to0$ extrapolated
data for $N_\tau=24$. In scheme II lattice data are ``corrected'' for
artefacts {\em before\/} integrating. The results clearly inherit the
bump at $\mu a\simeq0.45$ also manifest in
Fig.~\ref{fig:density-compare}, which we now believe to be an IR
artefact. However, this bump is absent (or strongly suppressed) in the
$j\to0$ limit, mirroring the absence of a significant bump in the
upper panel of Fig.~\ref{fig:density}.  This 
extrapolation reduces the ratio $p/p_{SB}$ from approximately 1.5 to
approximately one in the quarkyonic regime. By contrast the scheme 0 data
have the  ratio $p/p_{SB}$ substantially
exceeding unity in the large-$\mu$ regime above deconfinement, which
probably reflects the fact that UV artefacts are not being fully
corrected here. For this reason we now prefer scheme I, where
for the coldest lattice $p/p_{SB}$ has a plateau with value
$\approx1$ (after $j\to0$) in the suspected quarkyonic region, only rising to
$\approx2$ at large $\mu$. Again, therefore, we conclude that for low
$T$ there is a range of $\mu$  where thermodynamic quantities scale
approximately the same as free quarks; that the evidence for a peak
above onset matching the expectations of $\chi$PT has substantially
diminished; and that $p/p_{SB}$ rises above unity in the deconfined
regime. By $T=141$ MeV ($N_\tau=8$), however, the ratio rises monotonically and the
distinction between these different regimes is largely washed out.  It
is clear, however, that the full story will only emerge once the
continuum and thermodynamic limits are both taken with care.

\begin{figure}[thb]
\includegraphics*[width=\colw]{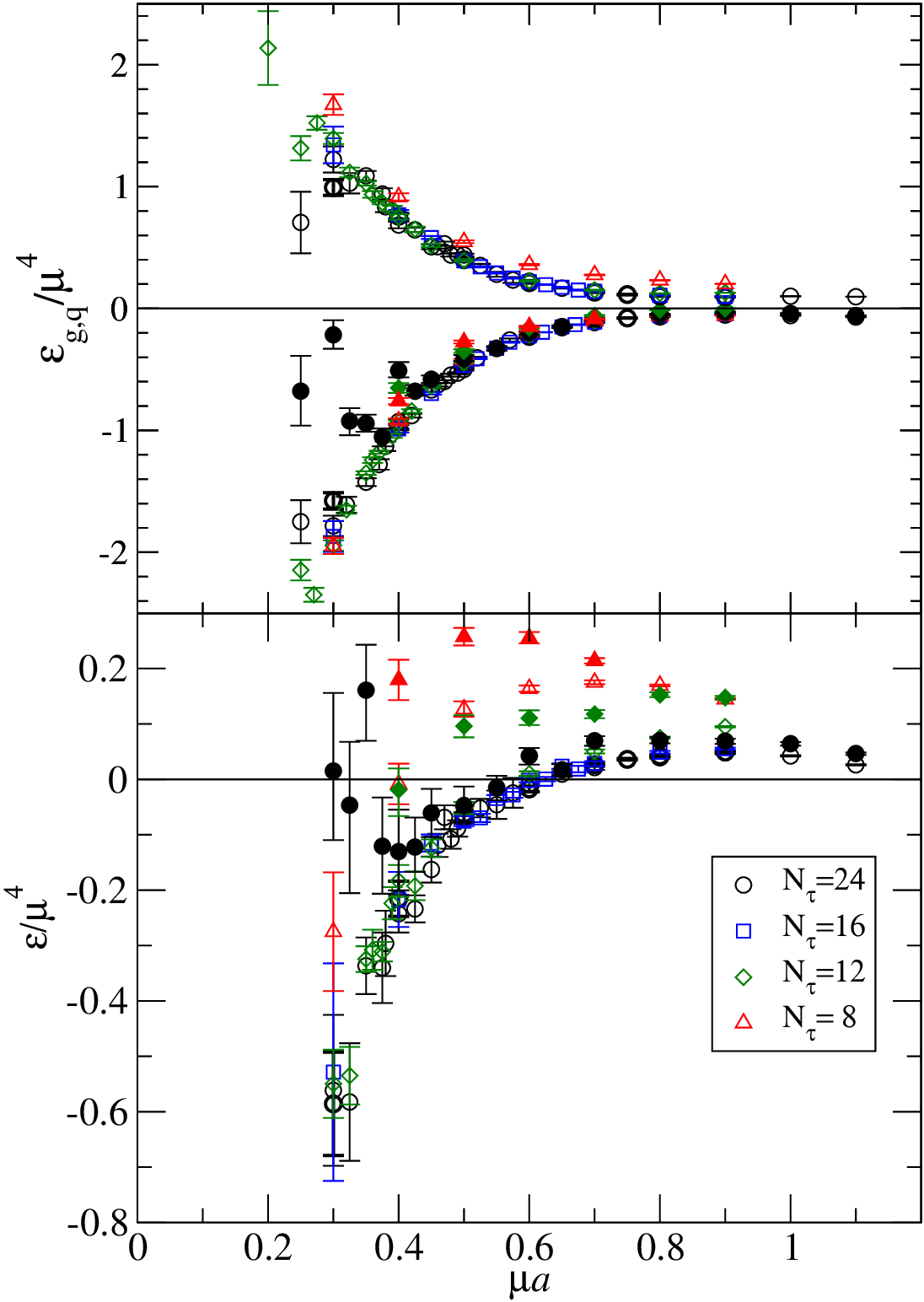}
\caption{Top: Renormalised quark (negative numbers) and gluon
    (positive numbers) energy density divided by $\mu^4$,
    at various temperatures, for $ja=0.04$ (open symbols) and
    extrapolated to $j=0$ (filled symbols).  Bottom: total energy
    density divided by $\mu^4$.}
\label{fig:energy}
\end{figure}

\begin{figure}[thb]
\includegraphics*[width=\colw]{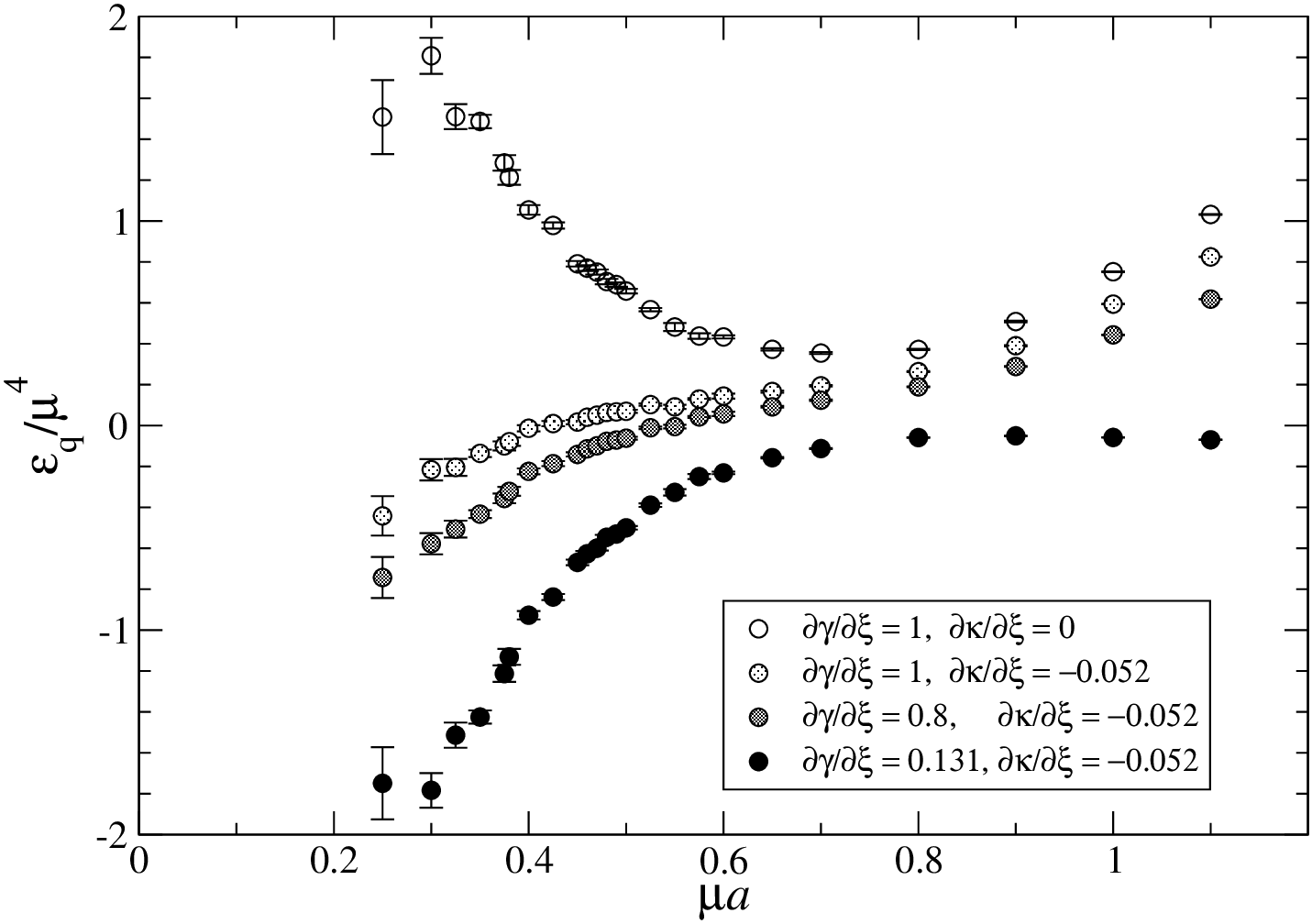}
\caption{Renormalised quark energy density divided by $\mu^4$
    density at $T=47$ MeV ($N_\tau=24$), $ja=0.04$, for different values of the
    Karsch coefficients $\partial\gamma_q/\partial\xi,
    \partial\kappa/\partial\xi$. }
\label{fig:quark-energy-compare}
\end{figure}

The quark and gluon contributions to the energy density, for
$ja=0.04$, are shown in the upper panel of
Fig.~\ref{fig:energy}.  We see that the quark energy density is
almost independent of temperature for all temperatures, while the
gluon energy density shows a clearly different behaviour only for the
highest temperature.  We find that the gluon energy density is
independent of the diquark source within errors, so these results are
representative for the $j\to0$ extrapolated data.  The quark
contribution is sensitive to the diquark source in the low-$\mu$
region, as can be seen from the $j\to0$ extrapolated data also shown
in Fig.~\ref{fig:energy}.

Comparing these results with the unrenormalised (and unextrapolated)
results in Figs 1 and 3 of Ref.~\cquarkyonic, we see a dramatic
difference.  Clearly, the proper renormalisation is crucial to any
reliable determination of the energy density, and in particular it is
clear that the terms proportional to $\partial\beta/\partial\xi$ and
$\partial\kappa/\partial\xi$ in \eqref{eq:epsG} and \eqref{eq:epsQ}
respectively cannot be ignored.  To illustrate this more clearly, we
show in Fig.~\ref{fig:quark-energy-compare} the quark contribution to
the energy density on the $12^3\times24$ lattice at $ja=0.04$,
computed using different values for the Karsch coefficients.  The open
circles correspond to the unrenormalised energy density which was
presented in Ref.~\cquarkyonic\ (note that the normalisation is
different).  The other data sets correspond to different values of
$\partial\gamma_q/\partial\xi$, with $\partial\kappa/\partial\xi$ set
to the value of $-0.052$ that was determined in Sec.~\ref{sec:karsch}.
We have chosen to use the tree-level value of 1, the value 0.131
determined in Sec.~\ref{sec:karsch}, and a value of 0.8, which is
similar to the value found for $\partial\gamma_g/\partial\xi$, and at
the margins of our 95\% confidence interval.  We see that using the
correct (non-zero) value for  $\partial\kappa/\partial\xi$ is most
important at low $\mu$, where this alone changes the sign of $\eq$.
At large $\mu$, the $\partial\gamma_q/\partial\xi$ term will dominate,
as it does at tree level.

It should be noted that the uncertainties in the Karsch coefficients
are not included in the total uncertainties in the plots shown here.
On the basis of Fig.~\ref{fig:quark-energy-compare} one may conclude
that these uncertainties will have an effect of $\order(100\%)$ in the
energy density.

Although $\eps_q$ appears to be negative at least for low $\mu$, and
possibly for all $\mu$-values considered here, the total energy
density $\eps=\eps_g+\eps_q$, shown in the bottom panel of
Fig.~\ref{fig:energy}, remains positive or consistent with zero
everywhere in the $j\to0$ limit.
Although on the face of it a negative value for $\varepsilon_q$ is
surprising,
it is notable that the renormalised quark energy density shown in
Fig.~\ref{fig:energy} has a qualitative resemblance to the
unrenormalised energy density \eqref{eq:epsQ}
measured for QC$_2$D with $N_f=4$ Wilson quark flavors in Fig.~5
of Ref.~\cite{Hands:2011ye}. The parameters used in that study
correspond to a much finer lattice, with $a/\sqrt{\sigma}$ having a
value approximately one-third that used here. It is conceivable,
therefore, that the Karsch coefficients for $N_f=4$ fall far closer to
their free-field values, and hence their neglect in
\cite{Hands:2011ye} is much better justified, reinforcing 
the conclusion that $\eq(\mu)<0$.

\begin{figure}[tb]
\includegraphics*[width=\colw]{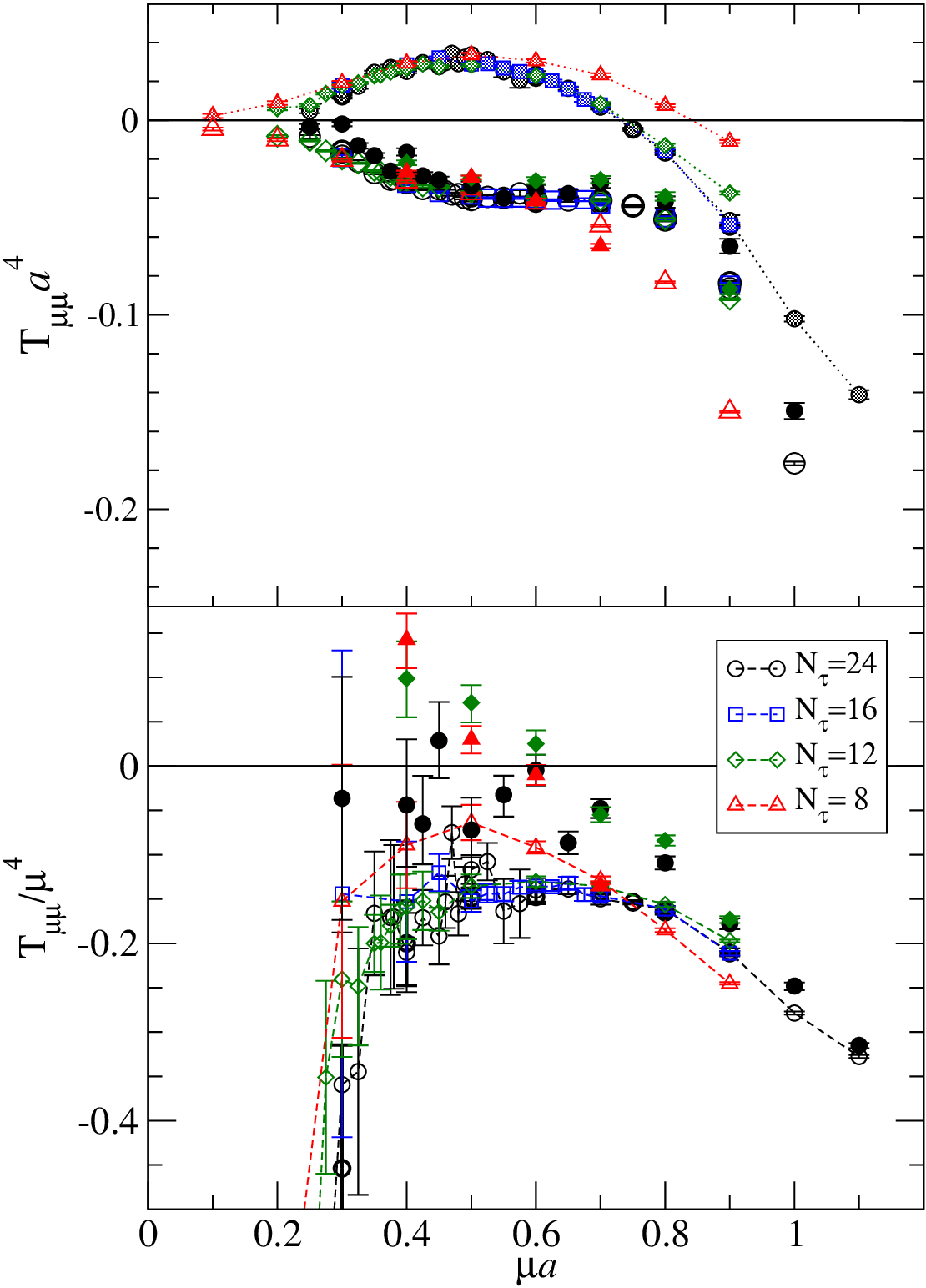}
\caption{Top: gluon (shaded symbols, dotted lines) and quark (open
  symbols) contributions to the trace anomaly, at $ja=0.04$.  The
  filled symbols denote the quark contributions extrapolated to $j=0$.
   Bottom: Total trace anomaly divided by $\mu^4$, for $ja=0.04$ (open
   symbols, dashed lines) and extrapolated to $j=0$ (filled symbols).}
\label{fig:trace-anomaly}
\end{figure}

Finally, we consider the trace anomaly, computed according to
Eqs~\eqref{eq:Tg}, \eqref{eq:Tq}, which is shown in
Fig.~\ref{fig:trace-anomaly}.  With the correct expression
\eqref{eq:Tq}, we now find the quark contribution to be negative for
all $\mu$, whereas in \cprev\ it had erroneously been presented as
positive.  Since the beta-functions only enter
into the expressions as overall constants, and our updated values
are not dramatically different from those used in Ref.~\cquarkyonic,
the qualitative behaviour of the $N_\tau=24, ja=0.04$ data is the
same as previously reported in Ref.~\cquarkyonic, apart from the
sign of the quark contribution.

For small and intermediate $\mu$, the gluon and quark contributions
have opposite signs and similar magnitudes, leading to a nearly
vanishing total trace anomaly in the region $0leq\mu a\lesssim0.7$.
The gluon contribution decreases for $\mu
\gtrsim0.5$ and becomes negative for $\mu a\gtrsim0.75$, while the
quark contribution has a plateau for $0.5\lesssim\mu a\lesssim0.75$
and increases rapidly in magnitude thereafter.  This leads to a
negative total trace anomaly at large $\mu$, which corresponds to the
positive and increasing pressure $p=(\eps-T_{\mu\mu})/3$ observed in
Fig.~\ref{fig:pressure}.

We see no difference in the trace anomaly between the two lowest
temperatures, $T=47$ and 70 MeV. At $T=94$ MeV and 141 MeV
($N_\tau=12$ and 8) the gluon contribution becomes larger (or less
negative) and the quark contribution becomes more
negative at large $\mu$.  The net effect of this, however, is to leave
the total trace anomaly nearly unchanged.

We find that the trace anomaly depends only weakly on the diquark
source for nearly all $T$ and $\mu$.  The main effect is to increase
the gluon contribution at large  $\mu$ and $T$, and to decrease the
magnitued of the quark contribution at low $T$, for large and small
$\mu$.  It is quite 
striking that there appears to be little or no dependence on either
temperature or diquark source in the region $\mu_o\lesssim\mu\lesssim
0.55/a$. 

Once again, it is instructive to compare with the $N_f=4$ study of
Ref.~\cite{Hands:2011ye}. In that case (see figs. 6 and 8 of
\cite{Hands:2011ye}), after taking into account the incorrect sign for
the quark contribution, the gluon
unrenormalised contribution to $T_{\mu\mu}$ is negative for all $\mu
\lesssim\mu_d$, while the quark contribution is positive, which is the
opposite of what we observe here.  However, this still leaves open the
possibility of the two contributions nearly cancelling, giving rise to
nearly-conformal matter in the quarkyonic region.

\subsection{Quark number susceptibility}
\label{sec:susc}

In a mathematical sense the Polyakov loop is a well-defined signal
for deconfinement, at least in pure gauge theories; physically it reveals
something about the behaviour of static color sources, which are well
approximated by heavy quarks, in a baryonic medium. 
Recent studies of a non-relativistic formulation of QC$_2$D~\cite{Hands:2012yy} took the
first step beyond the static approximation, and revealed a non-trivial $T$-
and $\mu$-depedence for $s$-wave states formed from heavy
quarks. Another observable related to confinement is
the quark number susceptibility $\chi_q\equiv\partial
n_q/\partial\mu$. This observable is usually thought of as
  encoding the fluctuations in the baryon (or quark) number,
  and is of particular interest as a measure of confinement or
  deconfinement of light quark degrees of freedom
  \cite{Aoki:2006br,Borsanyi:2010bp, Bazavov:2009zn, Hands:2010vw}.
  If quarks are  
  confined inside hadrons, the fluctuations of the quark number and
  hence the susceptibility will be suppressed, since increasing the
  quark number entails exciting a baryon, which requires a large
  amount of energy.  If quarks are not confined, it is possible to
  excite a single quark, which requires much less energy, giving a
  larger quark number susceptibility.

This link between $\chi_q$ and deconfinement is clear in the case of
QCD, where all baryons are heavy.  In the case of QC$_2$D the
situation is less clear, since the lightest baryons are the
pseudo-Goldstone diquarks, and large fluctuations are possible even in
the confined phase.  Nonetheless, it is of great interest to study
fluctuations in quark number at large density and low temperature.
The only previous such study is Ref.~\cite{He:2008zzb}, where the
Dyson--Schwinger equation in the rainbow approximation was employed.
Hence
QC$_2$D offers an opportunity for a first systematic non-perturbative
study of $\chi_q$ in this r\'egime.

For an ideal gas of massless (continuum) quarks and gluons, 
at temperature $T$ and chemical potential $\mu$, we have:
\begin{align}
n_{SB}^{\text{cont}}&= N_f N_c  \left( \frac{\mu T^2}{3} + \frac{\mu^3}{3 \pi^2} 
\right) \ , \label{eq:nq-ideal}
\\
\chi_{SB}^{\text{cont}}&= N_f N_c  \left( \frac{T^2}{3} + \frac{\mu^2}{\pi^2} \right) \ .
\label{nchinSB}
\end{align}
Now consider the quark action (\ref{eq:Slatt}) rewritten in the form
$\bar\Psi{\cal M}\Psi$, where we have introduced the bispinors
$\Psi\equiv(\psi_1, C^{-1}\tau_2\bar\psi_2^{tr})^{tr}$, 
$\bar\Psi\equiv(\bar\psi_1,-\psi_2^{tr}C\tau_2)$;
see Ref.~\cite{Giudice:2011zu} for details.
From the definition of $\chi_q$ we have:
\begin{widetext}
\begin{equation}
\chi_q=\del{n_q}{\mu}
=\frac{T}{V_s}
 \left\{-\bigg\bra\Big[-\bar\Psi\del{\mathcal{M}}{\mu}\Psi\Big]\bigg\ket^2
+ \bigg\bra \Big[ -\bar\Psi \del{\mathcal{M}}{\mu} \Psi \Big]^2 \bigg\ket
+ \bigg\bra\Big[ -\bar\Psi\del{^2\mathcal{M}}{\mu^2}\Psi\Big]\bigg\ket\right\}
 \label{chi3terms} \ .
\end{equation}
From this equation we can identify four different terms:
\begin{alignat}{2}
T_1&= - \bigg\bra\Big[-\bar\Psi\del{\mathcal{M}}{\mu}\Psi\Big]\bigg\ket^2 
 &&= - \bigg\bra\tr\Big[\mathcal{M}^{-1}\del{\mathcal{M}}{\mu}\Big]\bigg\ket^2\,,
 \label{T1} \\
T_2&= + \bigg\bra \Big[ -\bar\Psi \del{\mathcal{M}}{\mu}\Psi\Big]^2\bigg\ket_{disc}
 &&=\phantom{-}\bigg\bra\tr\Big[ \mathcal{M}^{-1}\del{\mathcal{M}}{\mu}\Big]
 \cdot\tr\Big[\mathcal{M}^{-1}\del{\mathcal{M}}{\mu}\Big]\bigg\ket\,,
 \label{T2} \\
C_1&= + \bigg\bra \Big[ -\bar\Psi \frac{\partial \mathcal{M}}{\partial \mu} 
\Psi \Big]^2 \bigg\ket_{conn}
 &&=- \bigg\bra \mbox{Tr} \Big[ \mathcal{M}^{-1}  
\del{\mathcal{M}}{\mu} \mathcal{M}^{-1}\del{\mathcal{M}}{\mu}\Big]\bigg\ket\,,
 \label{C1} \\
T_3&= + \bigg\bra\Big[-\bar\Psi\del{^2\mathcal{M}}{\mu^2}\Psi \Big]\bigg\ket
 &&=\phantom{-} \bigg\bra\tr\Big[\mathcal{M}^{-1}\del{^2\mathcal{M}}{\mu^2}\Big]\bigg\ket
 \label{T3}  \, .
\end{alignat}
\end{widetext}

The second term of Eq.~(\ref{chi3terms}) yields two terms, 
$T_2$ and $C_1$, because there are two ways to contract the spinors.

The calculation of the traces is done using unbiased estimators, introducing 
$N_\eta$ complex noise vectors $\eta$ with the properties: 
$\bra\eta_x\ket =0$ and $\bra\eta_x\eta_y\ket = \delta_{x y}$.
For example, the determination of the trace, used for $T_1$ and $T_2$, is based
on the following relation:
\begin{equation}
\tr \Big[\mathcal{M}^{-1}\del{\mathcal{M}}{\mu}\Big] =
\frac{1}{N_\eta}\sum 
\eta^*_{x \alpha i}\left( \del{\mathcal{M}}{\mu} 
\right)_{x \alpha i ; y \beta j} \!\!\!\!\!\!\!\!
\mathcal{M}^{-1}_{y \beta j ; z \gamma k}
\eta_{z \gamma k} \ .
\label{trmmdm}
\end{equation}
Because two independent source vectors are required to compute $T_2$,
we refer to this term as ``disconnected''; the other three
``connected'' terms need only one source vector.

It turns out that the connected term gives an important contribution
to $\chi_q$ at low and high values of the
chemical potential and therefore cannot be considered negligible. Moreover,
it changes sign around $\mu a\approx 0.66$. On the other hand, the terms 
$T_1$ and $T_2$ are equal within errors but with opposite sign, {\em
  i.e.,} their net contribution is consistent with zero everywhere,
except possibly around the onset transition.

\begin{figure}[ht]
\center
\vspace{5mm}
\includegraphics[width=\colw]{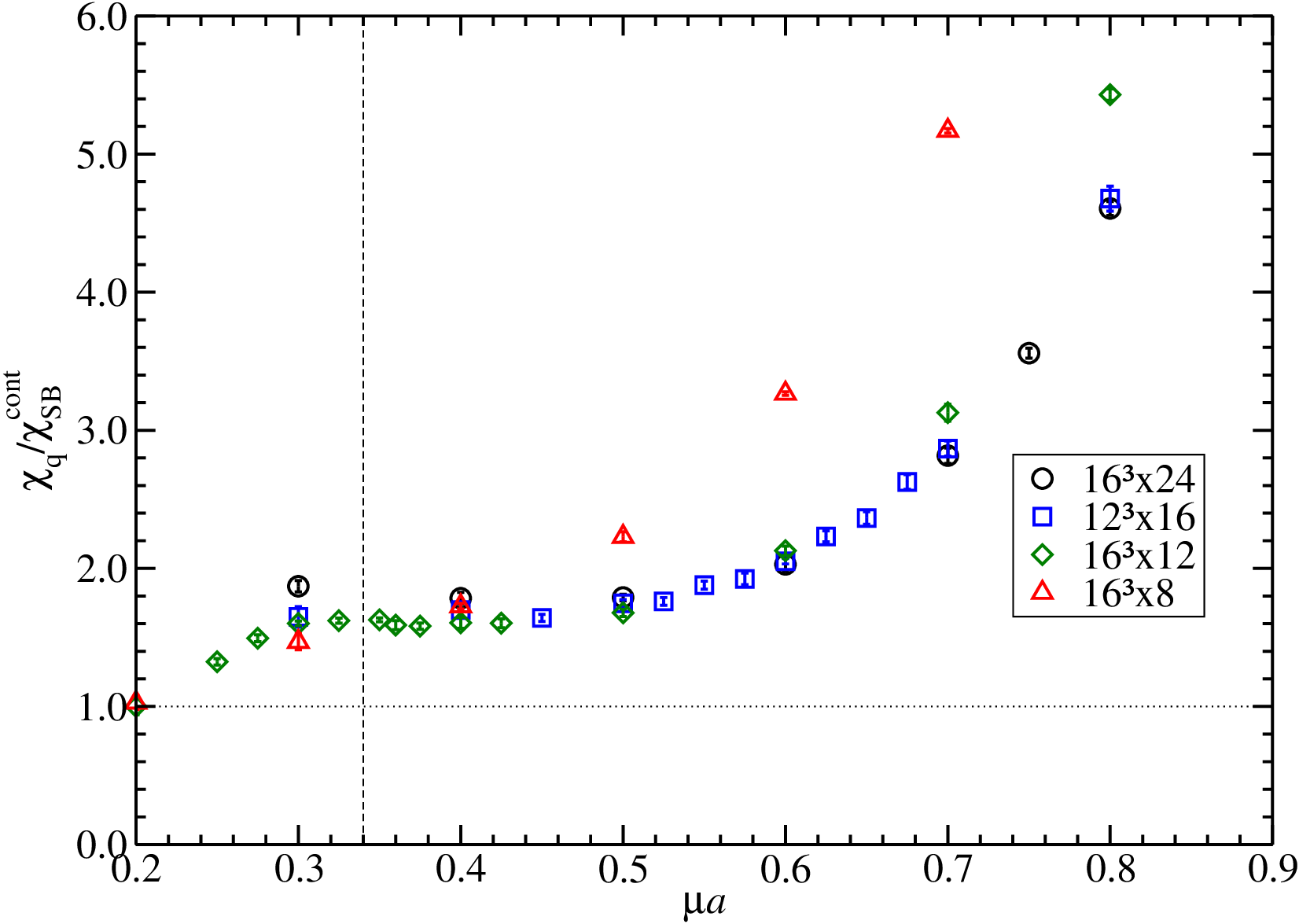}\\
\caption{Ratio $\chi_q/\chi_{SB}^{\text{cont}}$ versus $\mu$, for $ja=0.04$.
The vertical dashed line marks the position of $\mu_o$. 
}
 \label{fig:chibyideal}
\end{figure}

All the systematic issues discussed in Sec.~\ref{sec:technical},
regarding the normalisation of data with the same quantity
calculated for free quarks, are also relevant for $\chi_q$
In Fig.~\ref{fig:chibyideal} we plot the ratio
$\chi_q/\chi_{SB}^{\text{cont}}$, for four different 
temperatures, versus the chemical potential.  For an ideal gas of
quarks and gluons this ratio would be a constant, see Eq.~(\ref{nchinSB}), 
and we see that
an approximate  plateau is actually present for $a\mu \lesssim 0.55$, 
at least for the three lowest temperatures; 
after this value we can see a sharp increase of $\chi_q$. 
The value of the plateau is $\chi_q/\chi_{SB}\approx 1.6$ which
is higher than the ideal value of 1.0.
Moreover, it is evident from this plot that $\chi_q$ is $T$-independent 
at low temperature, since there is no significant deviation in the 
behaviour of the three curves.
This is to be contrasted  with the Polyakov loop in
Fig.~\ref{fig:polyakov}, which shows deconfinement 
for three different values of $\mu=\mu_d(T)$, as 
the temperature is varied.  Only for the highest temperature do we see
a different behavour signalling a different phase.

\begin{figure}[ht]
\center
\includegraphics*[width=\colw]{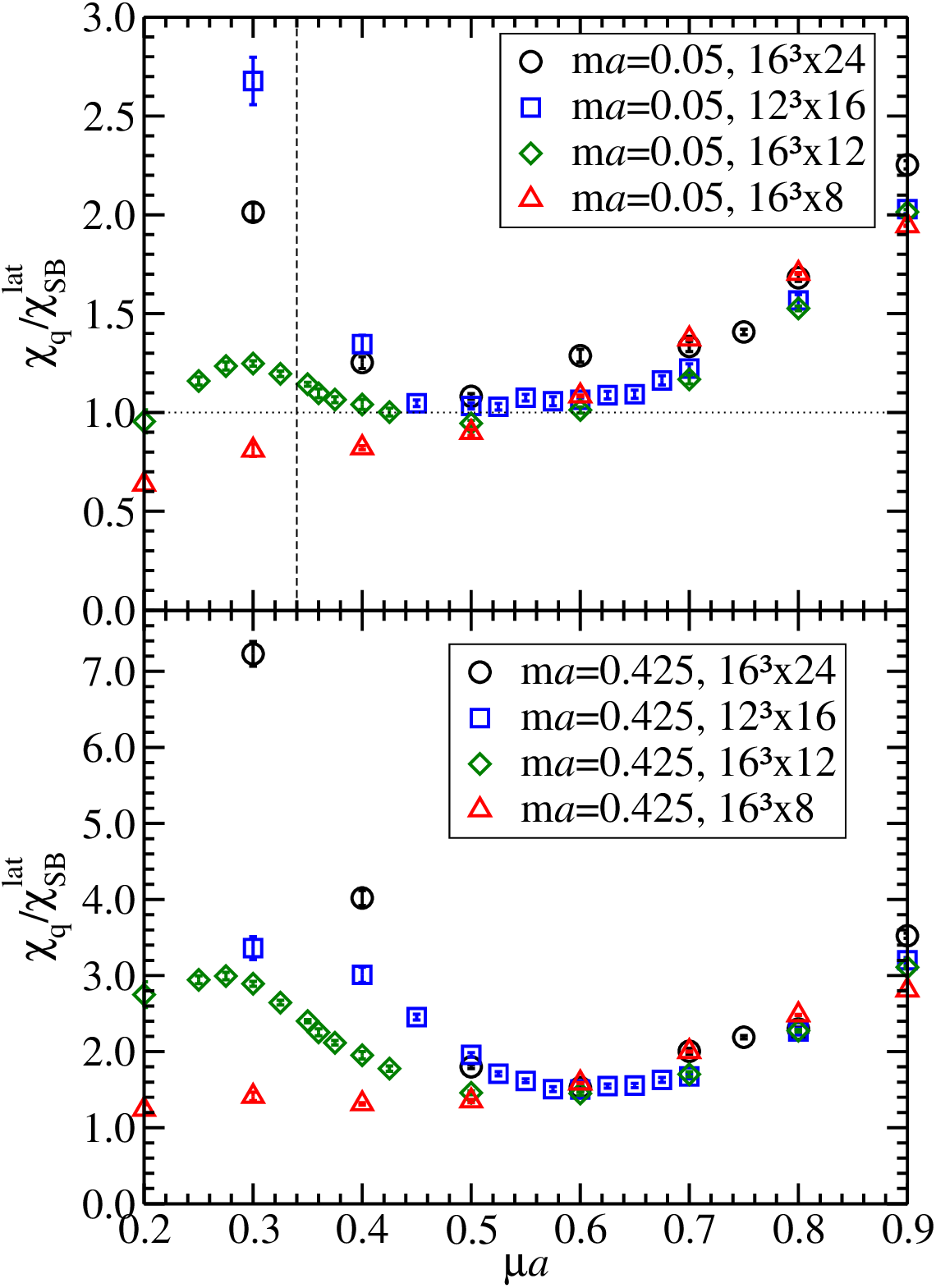}
\caption{The ratio between the measured quark number
susceptibility at $ja=0.04$ and the ideal value  
for \emph{lattice} free fermions for two values of the fermion mass: 
$m_q=0.05$ (top) and $m_c=0.42$ (bottom).
The vertical dashed lines mark the position of $\mu_o$.}
\label{fig:chibyLat}
\end{figure}

It is also instructive to compare the numerical results
with the equations corresponding to Eq.~(\ref{nchinSB}) but taking in account 
the finite volume and the lattice discretisation.
In Eq.(26) of Ref.~\cite{Hands:2006ve}, the expression for
the quark number density $n_{SB}^{\text{lat}}$ for free Wilson fermions
on the lattice is presented, from which $\chi_{SB}^{\text{lat}}$ is easily obtained.
Fig.~\ref{fig:chibyLat} plots the ratio $\chi_q/\chi_{SB}^{\text{lat}}$
for two values of the quark mass used in the determination of
$\chi_{SB}^{\text{lat}}$, the subtracted bare quark mass
  $m_q=1/2\kappa-1/2\kappa_c$ and the `constituent' quark mass
  $m_c=m_\rho/2$.
In this case we observe a different behaviour for $a\mu \lesssim 0.45$, 
but now in the quarkyonic regime there is a discernable plateau
with a ratio compatible with  one,
ie.\ the system is behaving as free fermions, with again 
an increase for higher values of $\mu$. 
Fig.~\ref{fig:chibyLat} demonstrates that the value of the mass used for the free 
fermions has a quantitative effect for this observable, in that the value of the
plateau is shifted when the mass is increased, but this does
not change the qualitative considerations.  The results using
  $m_q$ are almost identical those obtained setting $m=0$.
These plots again confirm the above scenario: we do not see any abrupt 
change for $\chi_q$ as a function of $T$,  whereas the Polyakov 
loop becomes different from zero at a $T$-dependent $\mu_d$.
Again, something different seems to emerge for the highest
  temperature, $T=141$ MeV ($N_\tau=8$).

\begin{figure}[ht]
\center
\includegraphics[width=\colw]{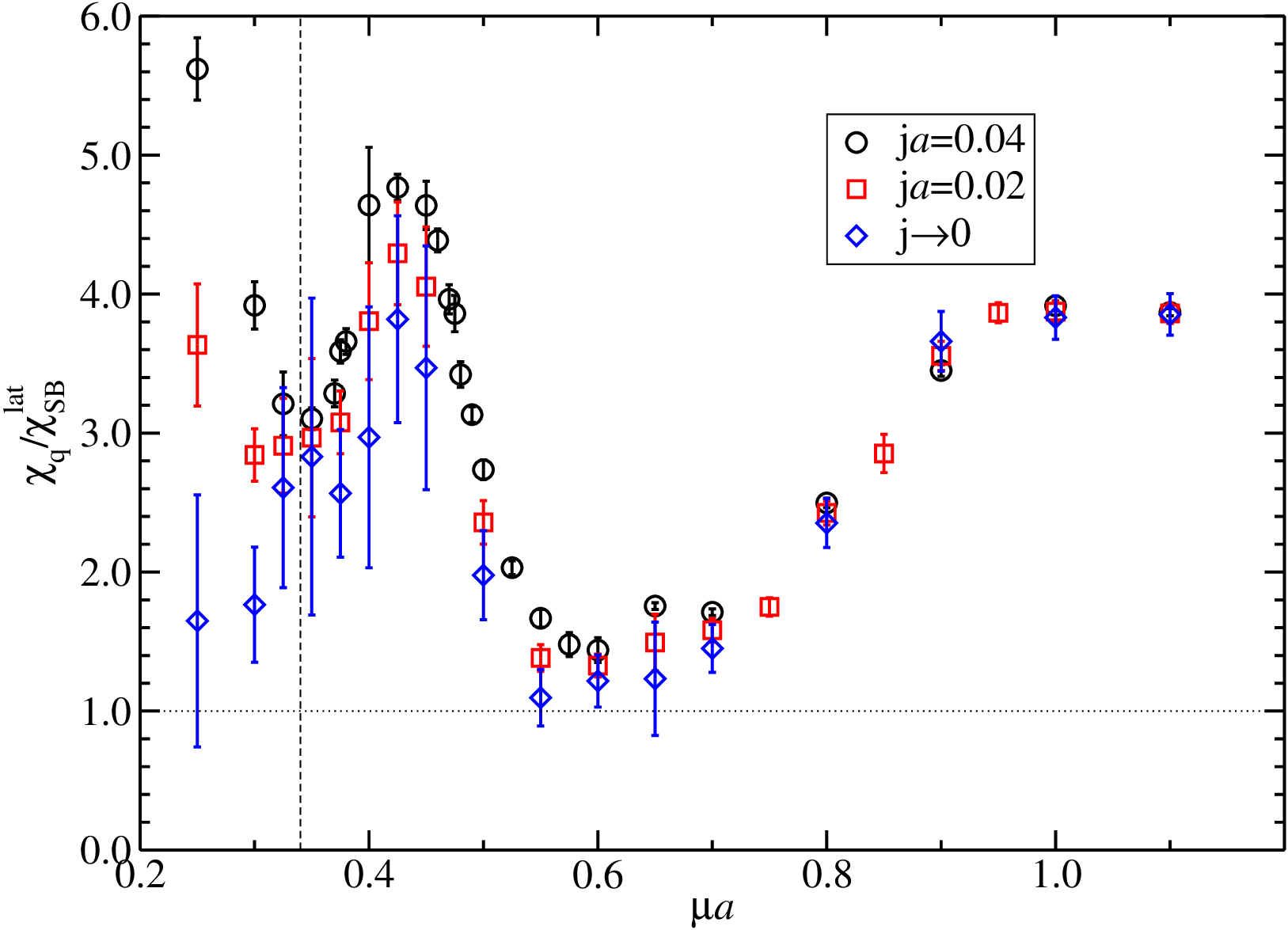}
\caption{The ratio between the measured quark number
susceptibility at different diquark sources $j$ and the ideal value  
for lattice free fermions with a fermion mass of $m_c=0.42$.
The vertical dashed lines mark the position of $\mu_o$.}
\label{fig:chi-j}
\end{figure}

The effect of the diquark source is illustrated in
Fig.~\ref{fig:chi-j}, where we show
$\chi_q/\chi_{SB}^{\text{lat}}(m_c)$ for the $12^3\times24$ lattice
and $ja=0.04, 0.02$ and 0.  We find that the diquark source only has
a significant effect for low $\mu$, where it increases the value of
$\chi_q$ slightly.

\subsection{A first look at chiral symmetry in the dense phase}
\label{sec:chisb}

An important issue which we have been hitherto unable to address is the
chiral properties of the ground state once $\mu\ge\mu_o$. 
This issue is of course of general theoretical interest when the phase diagram of any
non-abelian gauge theory is discussed; in the current context it is of
particular interest since the original
description of the quarkyonic phase in SU($N_c$) gauge theory was in terms of a
chirally
symmetric but confined medium, ie. one in which the chiral condensate
$\bra\bar\psi\psi\ket\to0$ as the bare quark mass $m\to0$~\cite{
McLerran:2007qj}. Later this picture was modified; chiral symmetry breaking via
a translationally non-invariant ``chiral spiral'' was postulated in
\cite{Kojo:2009ha}.
For theories of the class exemplified by QC$_2$D where the
relevant mass scale is set by $m_\pi$, chiral symmetry is necessarily always
broken explicitly by a bare quark mass $m$; in this case the question is how the
condensate $\qq$ scales with $m$ as $m\to0$.

It is clearly desirable to determine the fate of chiral symmetry
breaking for the case of QC$_2$D by a lattice calculation. Indeed,
$\bra\bar\psi\psi\ket$ was examined in early studies such as
\cite{Hands:2000ei} using staggered lattice fermions, and reasonable
quantitative agreement found over a decade of quark mass with the
prediction of leading order $\chi$PT for $T\to0$, namely that for
$\mu<\mu_o$ the chiral condensate is $\mu$-independent, and for
$\mu\ge\mu_o$
\begin{equation}
\bra\bar\psi\psi\ket\propto\frac{m}{\mu^2}.
\label{eq:chicondXPT}
\end{equation}
Unfortunately, since the global symmetries of staggered fermions do
not coincide with those of continuum QC$_2$D~\cite{Hands:2000ei},
these results are not directly applicable. In any case, no attempt was
made to explore beyond the r{\'e}gime of applicability of $\chi$PT.


However, our use of Wilson fermions precludes any direct study in the current
simulation, since this formulation violates chiral symmetry explicitly.
Our strategy therefore is to calculate a chiral order parameter using a fermion
formulation with manifest chiral and baryon number symmetries
using  the gauge backgrounds ensembles generated with Wilson quarks. The
disparity
between valence and sea quarks violates unitarity; we mitigate this
uncontrolled approximation by tuning the mass of the valence quarks so that the
pion mass coincides with that used in the simulation; once
$\mu\not=0$ the onset transition of the valence quarks should then at least
coincide with the true value.

Rather than the obvious choice of staggered fermions for the valence
quarks, we found it expedient to use the existing code for $N_f=2$ Wilson
fermions
with the parameter $r$ (which has the conventional value of unity in
(\ref{eq:Mwils}))
set to zero. For $j=0$ this is equivalent to eight identical staggered fermions
with mass $m=(2\kappa)^{-1}$. For non-zero lattice spacing and $\mu\not=0$ the
action has a U(8)$\otimes$U(8) global symmetry which is broken by $m\not=0$
(explicitly) or $\bra\bar\psi\psi\ket\not=0$ (spontaneously) to U(8)$_V$
(the subscript denotes vectorlike), which incorporates the U(1)$_B$ of
baryon number. A diquark source $j\not=0$ breaks
U(8)$_V$ to a SU(2)$\otimes$SU(2) which preserves isospin but no longer includes
U(1)$_B$.

\begin{figure}[tb]
\center
\vspace{5mm}
\includegraphics[width=\colw]{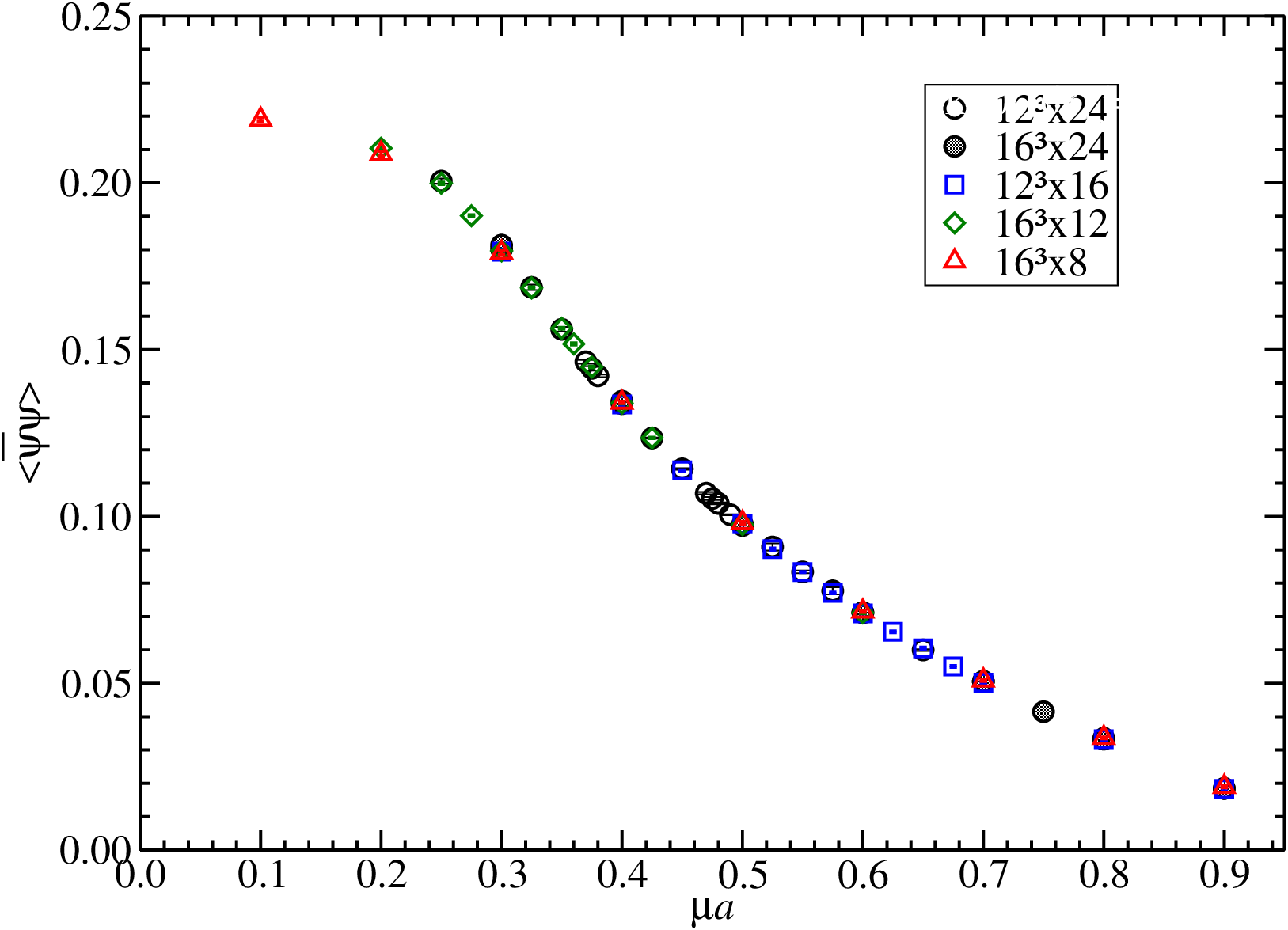}
\caption{$\bra\bar\psi\psi\ket$ versus $\mu$ for $r=0$, $ja=0.04$
and $\kappa=8.0$.}
\label{plot-chiral-tot}
\end{figure}
By studying effective mass plots as the valence $\kappa_V$ was varied we found
that $\kappa_V=8.0$ gave the closest match to the value $m_\pi a=0.66(2)$ found
for $\beta=1.9$, $\kappa=0.168$. Fig.~\ref{plot-chiral-tot} then shows the
resulting chiral condensate as a function of $\mu$ for the various lattices
studied. Note that $ja=0.04$ throughout, since this was found to yield a less
noisy and more stable signal -- hence these results are
not reproducible using pure staggered fermions. Two things are apparent; first
the shape of the curve is in qualitative agreement with the old staggered
results of \cite{Hands:2000ei} over the whole range of $\mu$ studied,
and thus consistent with (\ref{eq:chicondXPT})
assuming an onset $\mu_oa\simeq0.3$.  Secondly, the results are
independent of temperature even up to $T=141$MeV ($N_\tau=8$).
It is also apparent that volume effects are negligible.

\begin{figure}[tb]
\center
\vspace{5mm}
\includegraphics[width=\colw]{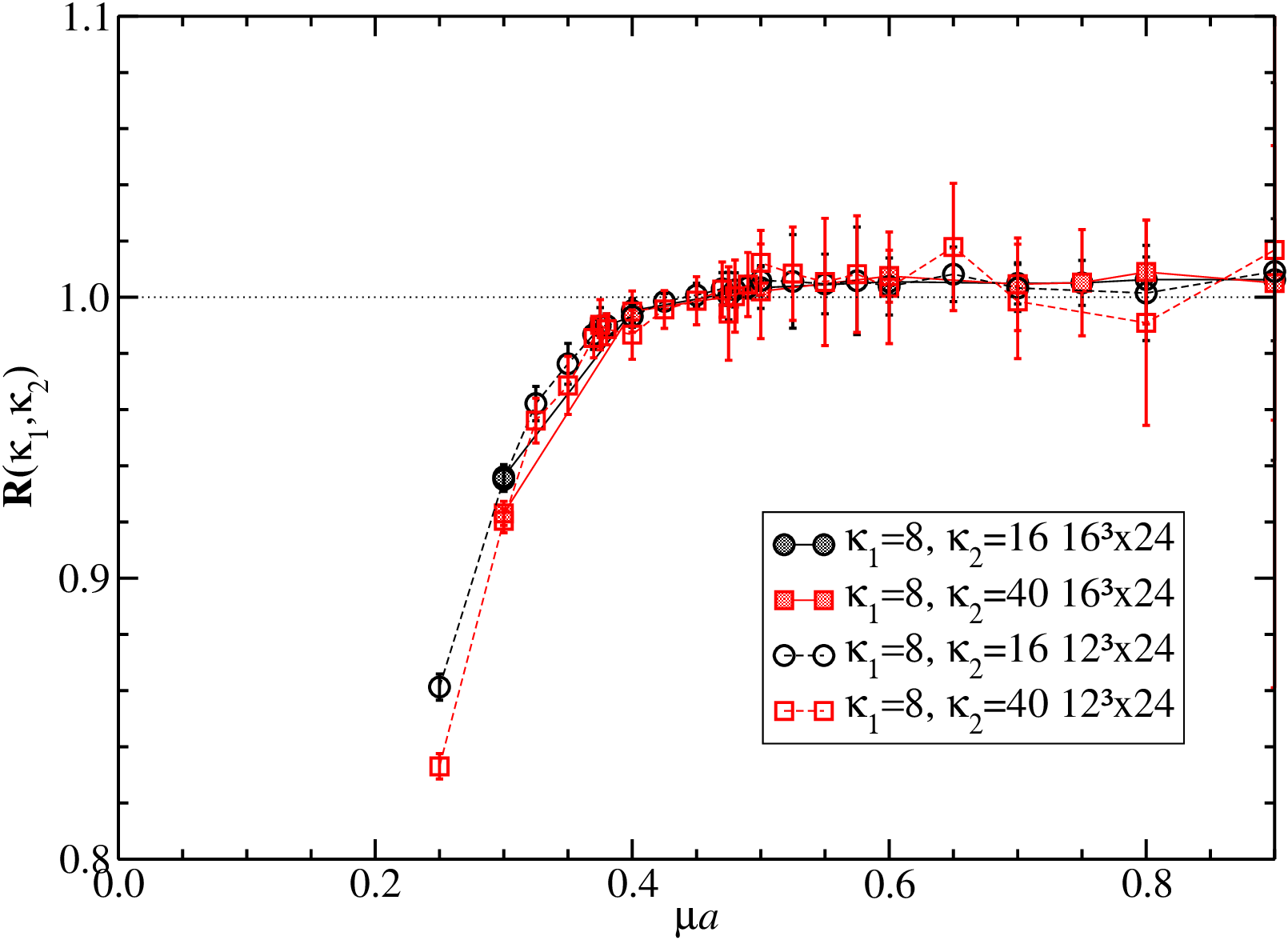}
\caption{The ratio
  $R(\kappa_1,\kappa_2)=[\kappa_1^2\pbp_1]/[\kappa_2^2\pbp_2]$ versus $\mu$
for $\kappa_0=8.0$ and $\kappa_1=16.0,40.0$, on the $12^3\times24$ and
$16^3\times24$ lattices with $ja=0.04$.}
\label{chiral-ratio}
\end{figure}

It appears that the chiral symmetry properties of the dense phase are
well-described by $\chi$PT.
In a sense the issuse of ``chiral symmetry restoration'' in QC$_2$D is academic,
since the onset scale is set on the assumption that chiral symmetry is
explicitly broken. Nonetheless, we can characterise the dense phase
by whether $\lim_{m_V\to0}\bra\bar\psi\psi(m_V)\ket$ vanishes or not.
We determine this by using three different values $\kappa_V=8$, 16 and 40, and
observing that with the field normalisations implicit in
(\ref{eq:Mwils}),
\begin{equation}
\frac{\kappa_{1}^2\bra\bar\psi\psi\ket_1}
{\kappa_{2}^2\bra\bar\psi\psi\ket_2}
 =\frac{m_{2}\bra\bar qq\ket_1}{m_{1}\bra\bar qq\ket_2}
 \begin{cases}=1&\pbp_0=0;\\
<1&\pbp_0\neq0,\; m_{2}<m_{1}.
\end{cases}
\end{equation}
Here $\bar q q$ denotes the scalar quark bilinear with conventional
normalisation and $\pbp_0$ is the chiral condensate in the
massless limit.
Fig.~\ref{chiral-ratio} shows this ratio plotted for both (8,16) and
(8,40) valence mass pairs on the $12^3\times24$ and $16^3\times24$ lattices as a
function of $\mu$, and clearly indicates
symmetry restoration for $\mu\gtrsim\mu_o$. Very similar plots are found for
the other temperatures explored. We therefore conclude that the gauge
field backgrounds at high baryon density in QC$_2$D are consistent with
chiral symmetry being unbroken by a scalar condensate, although the exotic
translationally-non-invariant scenario of \cite{Kojo:2009ha} is not ruled out.

We find no significant difference between our results for the
  $12^3$ and $16^3$ lattices.  This suggests that the chiral order
  parameter responds smoothly as 
$m$ increases, with no indication at this stage of a phase transition (indeed
the results are compatible with the predictions of chiral perturbation theory).
However, in the absence of any systematic finite volume scaling study, and in
light of the uncontrolled systematic uncertainties involved in our use of
different actions and quark masses for sea and valence quarks, this should,
like all the other results in this section, be taken as merely indicative.

\section{Conclusions and outlook}
\label{sec:conclude}

We have carried out the first extensive exploration of the phase
diagram of two-color QCD (QC$_2$D) in the $(T,\mu)$ plane using
first-principles lattice simulations.  Our main findings are
summarised in the tentative phase diagram of Fig.~\ref{fig:phases}.
We find evidence of three distinct regions:
\begin{enumerate}
\item A vacuum/hadronic phase, with $\qq=0, \braket{L}\approx0, \pbp\neq0,
  n_q\approx0$, at low $T$ and $\mu\lesssim\mu_0=m_\pi/2$;
\item A quarkyonic phase at low $T$ and intermediate to large $\mu$,
  which is confined ($\braket{L}\approx0$) and
  characterised by a chiral condensate which vanishes in the chiral limit,
  Stefan--Boltzmann scaling of bulk thermodynamic 
  quantities (including a nearly vanishing trace anomaly) and BCS
  scaling of the diquark condensate;
\item A deconfined quark--gluon plasma phase at high $T$ (and/or large
  $\mu$).
\end{enumerate}

The main difference from our previous studies is that the BEC region
has disappeared as a consequence of the $j\to0$ extrapolation and a
better understanding of the volume dependence and appropriate
normalisation of our results.  The BEC window would be expected to
reappear for smaller $m_\pi/m_\rho$.

While we have clearly defined the finite-temperature deconfinement
transition at $\mu=0$, and find clear evidence of a deconfinement
temperature that decreases as $\mu$ increases, the exact nature and
location of this transition at large $\mu$ remain elusive.  In order
to pin down this transition, and also to precisely locate the
superfluid-to-normal transition, we need to perform fine temperature
scans by varying $N_\tau$ at fixed chemical potential.  This is
currently underway.  We are also studying the static quark potential,
which should give further insight into the nature of this transition.

For the first time in this paper we have attempted to calculate
renormalised energy densities via an estimate of Karsch coefficients
obtained from simulations on anisotropic lattices. We find the
resulting corrections to our earlier results are substantial, and
indeed strongly suggest the quark contribution $\eq(\mu)$
is negative, implying that the physical requirement $\eq+\eg>0$
arises from a cancellation between terms of opposite
sign. Considerably greater accuracy will be required, therefore,
before we can contemplate {\it eg.} using lattice results as input for
the solution of the Tolman--Oppenheimer--Volkoff  equations used in
modelling relativistic stars.

We find that the quark number susceptibility $\chi_q$ is remarkably
independent of the temperature up to $T\simeq100$MeV, and stays close
to its noninteracting value in the quarkyonic region.  Most
strikingly, it shows little if any sensitivity to the deconfinement
transition, which occurs at different chemical potentials for our 4
temperature values.

The observation that $\chi_q$ is not a proxy for the Polyakov loop $L$
in the r\'egime of high quark number density suggests the following
conjecture. In the quarkyonic region $\mu_o<\mu<\mu_d$ the bulk
observables $p$, $n_q$, $\chi_q$ are approximately equal to the
free-field values $p_{SB}$, $n_{SB}$ and $\chi_{SB}$. This is
indicative of weakly self-bound quark matter, ie.\ with
$E_F=\mu\approx k_F$. The transition at $\mu=\mu_d$ (which coincides
with deconfinement as signalled by $L\neq0$ only in the limit $T\to0$)
is to a more strongly self-bound r\'egime with $E_F<k_F$. Since the
quarkyonic phase is confining, we interpret weak self-binding as the
quarks interacting via {\em binary\/} short-ranged interactions. For
$\mu> \mu_d$, the interaction is screened and may not be much
longer-ranged, but in this case deconfined quarks may interact with
several other quarks in the vicinity leading to stronger
binding. These considerations are related to interactions within bulk
quark matter, involving quarks with all energies less than the Fermi
energy, which hence are not sensitive to temperature $T$.

By contrast, the observed temperature-sensitivity of the Polyakov loop
$L$ (see Fig.~\ref{fig:polyakov}) suggests that in this case the
relevant physics is associated with degrees of freedom close to the
Fermi surface, which are readily thermally excited. These, of course,
are the same degrees of freedom relevant for transport. Our results
contrast with the findings of analytic and numerical studies of
QCD-like theories at weak coupling in small volumes of characteristic
scale $R\ll \Lambda_{QCD}^{-1}$ \cite{Hands:2010vw,Hands:2010zp}, and
a recent study of cold dense QCD with heavy quarks
\cite{Fromm:2012eb}, both of which show a coincidence in the rise of
$L$ and $\chi_q$. This suggests that a full description of
deconfinement at high baryon density requires a thermodynamic limit
and light, mobile degrees of freedom.

The main shortcoming of this study is that it has been performed with
a single, relatively coarse lattice spacing.  Although, as observed
in Ref.~\cquarkyonic, the main results are in qualitative agreement
with the earlier results \cdeconf\ obtained on a coarser lattice with
$a=0.23$fm, we also observe significant quantitative discrepancies,
and substantial lattice artefacts for $\mu a\gtrsim0.75$.  To get this
under control it will be necessary to repeat our simulations on a
finer lattice.  Thanks to the extensive investigation of parameter
space reported in Sec.~\ref{sec:vacuum}, we are in a good position to
carry this out, and these simulation are underway.

The large quark mass is another source of systematic uncertainty;
moreover our discussion of chiral symmetry in Sec.~\ref{sec:chisb}
is at best exploratory, and must in due course be supplemented by a calculation
respecting unitarity.
It is clear that QC$_2$D must be treated as a separate theory
and cannot be viewed as an approximation to QCD -- indeed, the
differences between the two theories become most stark in the chiral
limit -- and there is hence no need to attempt to match quark masses
to those in the real world.  Still, many analytical results have been
obtained in or near the chiral limit.  Also, as already mentioned, we
would expect a BEC region to open up near $\mu_o$ for smaller values
of $m_\pi/m_\rho$, and a fuller understanding of the BEC--BCS
crossover would be valuable.  For all these reasons, simulations with
smaller quark masses would be of great interest, and such simulations
are underway.

In addition to the quantities considered here, we are in the process
of computing the Landau-gauge gluon and quark propagators.  This will
allow us to check the assumptions involved in model solutions of the
superfluid or superconducting gap equation, and may form a direct link
with functional methods such as the functional renormalisation group
and Dyson--Schwinger equations.  These do not suffer from the sign
problem, but rely on assumptions regarding the form of propagators and
higher order vertices.  This will be addressed in a forthcoming
publication.

\begin{acknowledgments}
This work is carried out as part of the UKQCD collaboration and the
DiRAC Facility jointly funded by STFC, the Large Facilities Capital
Fund of BIS and Swansea University.  We thank the DEISA Consortium
(www.deisa.eu), funded through the EU FP7 project RI-222919, for
support within the DEISA Extreme Computing Initiative.  The simulation
code was adapted with the help of Edinburgh Parallel Computing Centre
funded by a Software Development Grant from EPSRC. JIS and SC
acknowledge the support of Science Foundation Ireland grants
08-RFP-PHY1462, 11-RFP.1-PHY3193 and 11-RFP.1-PHY3193-STTF-1.
JIS acknowledges the support and hospitality of the Institute for
Nuclear Theory at the University of Washington, where part of this
work was carried out. We warmly thank Joyce Myers and Seyong Kim for
their help.
\end{acknowledgments}

\bibliography{density,hot,lattice}

\end{document}